\documentclass[10pt,twocolumn]{article}

\usepackage{amsmath,amsfonts,amssymb,bm}
\usepackage{mathtools}
\usepackage{physics}

\usepackage{graphicx}
\usepackage[dvipsnames,svgnames,x11names]{xcolor}

\usepackage{geometry}
\usepackage{newtxtext,newtxmath} % Times-like text & math
\usepackage{anyfontsize} % for error of Font shape `OT1/cmr/bx/n' in size <10.5> not available
\usepackage{inconsolata}

\usepackage[labelfont=bf]{caption}
\usepackage{subcaption}

\usepackage{titlesec}   % section formatting
\usepackage{nomencl}    % nomenclature
\usepackage{float}      % improved float control
\usepackage{lscape}     % landscape pages
\usepackage{nopageno}   % no page numbers
\usepackage{ulem}       % underline the references!
\usepackage{nicefrac}
\usepackage{adjustbox}
\usepackage[english]{babel} % hypenthation

\usepackage[comma,authoryear,round,sort]{natbib}
\usepackage{multibib}
\newcites{discussion}{Discussion References}

\usepackage{hyperref}
\hypersetup{
	unicode=true, 
	pdftitle={Sharp-interface Simulations of Energetic Multiphase Flows with Large Density and Viscosity Ratios},
	pdfauthor={Tzu-Yao Huang},
	pdfnewwindow=true, 
	colorlinks=true, 	% (false,true)
	pdfborder={0 0 0},	
	linkcolor=black,
	linktoc=all, 		% (none,all) 
	citecolor=black,
	urlcolor=blue,
	breaklinks=true,
}

\usepackage[capitalise,nameinlink,noabbrev]{cleveref}
\creflabelformat{equation}{#2#1#3}

\usepackage{lipsum}
\makeatletter

\renewcommand\@biblabel[1]{}

\titleformat{\section}
  {\normalfont\bfseries\MakeUppercase}
  {\thesection}
  {1em}
  {}

\titleformat{\subsection}
  {\bfseries\normalsize\color{darkgray}}
  {\thesubsection}
  {1em}
  {}

\titleformat{\subsubsection}
  {\itshape\normalsize}
  {\thesubsubsection}
  {1em}
  {\uline}

\titlespacing{\section}{0pt}{1.5\baselineskip}{0.5\baselineskip}
\titlespacing{\subsection}{0pt}{1.0\baselineskip}{0.25\baselineskip}
\titlespacing{\subsubsection}{0pt}{0.5\baselineskip}{0.125\baselineskip}

\makeatother

\usepackage{fancyhdr}
\fancypagestyle{firstpage}
{
    \fancyhead[L]{}    
    \fancyhead[R]{\fontsize{10pt}{12pt}\selectfont{}36\textsuperscript{th} Symposium on Naval Hydrodynamics\\Busan, Korea, June 14 -- 19, 2026}
}

\title{\LARGE{}\bfseries{}Sharp-interface Simulations of Energetic Multiphase Flows with Large Density and Viscosity Ratios}
\date{}

\usepackage[noblocks]{authblk}
\author{\Large{}%
T.-Y.~Huang\textsuperscript{1}, 
N.~Valle\textsuperscript{1}, 
A.~K.~Lidtke\textsuperscript{2}, 
K.~Hendrickson\textsuperscript{3}, 
G.~D.~Weymouth\textsuperscript{1}\protect\\[0em]
(\textsuperscript{1}Delft University of Technology, the Netherlands, 
\textsuperscript{2}Maritime Research Institute Netherlands (MARIN), the Netherlands, 
\textsuperscript{3}University of North Carolina at Chapel Hill, United States)%
}

\usepackage{titling}
\pretitle{\begin{center}\LARGE}    % format title
\posttitle{\end{center}} % space after title
\preauthor{\vspace{-1em}\begin{center}\large}
\postauthor{\end{center}\vspace{-0.75em}} % <--- space after authors
\predate{} \postdate{}             % no date

\usepackage{hanging}

\newcommand{\ie}{\textit{i.e.}~}

\newcommand{\nfot}{\nicefrac{1}{2}}

\newif\ifshowchanges
\showchangestrue               % ← set \showchangesfalse to hide all marks

\colorlet{addcolor}{teal!70!black}
\colorlet{delcolor}{red!65!black}
\colorlet{notecolor}{blue!60!black}

\begin{document}
% The following commands replace \sloppy
% \tolerance 1414
% \hbadness 1414
% \emergencystretch 1.5em
% \hfuzz 0.3pt
% \widowpenalty=10000
% \vfuzz \hfuzz
% \raggedbottom

% \sloppy

\pagenumbering{gobble} % Suppress pagenumber

\setcounter{secnumdepth}{-1}

\maketitle

\thispagestyle{firstpage}

% \begin{multicols*}{2}

\section{Abstract}
Flows with high density ratios, such as wave breaking and air entrainment in maritime applications, remain challenging to simulate due to their energetic and strongly nonlinear nature. 
In such regimes, maintaining numerical robustness is difficult when using the commonly adopted velocity-based formulation. 
The Consistent Mass-Momentum (CMOM) transport framework improves numerical robustness by enforcing fundamental physical properties, most notably momentum conservation and semi-discrete energy-conserving. 
However, CMOM replaces the advection of a continuous velocity field with that of a discontinuous momentum field. 
When combined with sharp interface methods, this leads to severe momentum shocks, for which conventional shock-capturing schemes are ineffective.
To reconcile physical fidelity with numerical robustness, this work proposes a Synchronized Donor-Region of Momentum fluxes (SynDRoM) that enforces monotonicity of the transported velocity field. 
The resulting algorithm effectively eliminates spurious velocity oscillations without sacrificing physical fidelity, as demonstrated through scalar transport and interfacial shear instability test cases. 
Beyond difficulties from large density ratio, improper estimation of viscosity in the vicinity of the interface can introduce numerical instabilities at finite time steps, thereby undermining overall robustness. 
To address this issue, a viscosity limiter based on the bounded kinetic viscosity concept is introduced and validated using a gravity-driven plane shear flow. 
Finally, a breaking wave simulation is performed to assess the combined performance of the proposed physics-preserving numerical schemes for multiphase flows.

\section{Introduction}

\subsection{Advection schemes for extreme density ratios}
Violent, turbulent multiphase flows are ubiquitous in naval hydrodynamics, arising in phenomena such as bow-wave breaking, wake air entrainment, propeller cavitation, and flows around appendages operating near the free surface. 
Since the earliest numerical simulations of multiphase flows \citep{1965Harlow}, the development of robust, efficient, and physically consistent numerical methods has remained a central challenge. 
In maritime applications, this challenge is exacerbated by extreme density ratios \citep{2010Desjardins_HighDensity}: approximately $10^3$ for air-water systems and up to $10^4$ for cavitating flows.

As a natural extension of single-phase solvers, non-conservative velocity formulations combined with interface-capturing techniques are widely used in multiphase simulations \citep{2009Dommermuth_CompHydro,2013Stern_Compship}. 
% Applying shock-capturing schemes, such as slope limiters, seems able to achieve the stability just as what they do in the single-phase flows. 
However, the inconsistent advection of velocity and density leads to misalignment between pressure gradients and density discontinuities across interfaces. 
This inconsistency can trigger numerical divergence, particularly when large pressure gradients are divided by small densities \citep{2021Jin_IBVOF,2021Arrufat_MassMomentumConserv}. 
Moreover, velocity-based formulations do not enforce momentum conservation or kinetic energy boundedness, allowing spurious transfers of momentum and energy directly with the advection terms \citep{2010Desjardins_HighDensity,2013Ghods_CRMT}.

Various stabilization strategies have been proposed to alleviate these issues, including the Ghost Fluid Method \citep{1999Fedkiw_GhostFluid,2020Peltonen_GFFoam}, Favre-averaged formulations \citep{2010Fu_Favre,2021Jin_IBVOF}, and artificially smoothed density profiles \citep{2009Fuster_ReviewBubbleSimulation}. 
While these approaches may improve numerical stability, they often compromise key physical properties such as momentum conservation, energy boundedness, or interface sharpness, and remain vulnerable to instabilities at high density ratios.

Consistent Mass-Momentum (CMOM) transport schemes, originally proposed by \citet{1998Rudman}, were specifically designed to address these challenges. 
By employing a conservative momentum formulation and synchronizing part of the momentum flux with the mass flux, CMOM schemes ensure consistent transport of discontinuities across flow variables. 
As a result, momentum conservation is satisfied by construction, which \citet{2021Mirjalili_EConserv,2023Valle_EConserv} find to be a prerequisite for semi-discrete energy conservation. 
Furthermore, through the synchronizing the fluxes, CMOM also avoids unbounded pressure-density division problem \citep{2023Valle_EConserv}. 
On the other hand, the computational overhead introduced by CMOM is small: momentum transport reuses the mass fluxes already computed during interface advection, while the additional conversion between velocity and momentum implies simple point-wise multiplication and division, which incurs only a negligible cost. 
Additionally, even when the pressure Poisson solve remains the dominant computational expense, its cost also decreases in practice because the improved consistency between mass and momentum transport produces a more regular flow field than traditional formulations.
These properties have made CMOM increasingly attractive for high-density-ratio multiphase flows \citep{2021Mirjalili_EConserv,2023Valle_EConserv,2023vanderEijk_FSIVOF,2023Kuhn_MMC,2024Li_JetRANS}.

Despite these advantages, CMOM combined with sharp-interface methods often produces oscillatory velocity fields due to momentum shocks. 
For incompressible flows, a satisfactory CMOM scheme should additionally ensure total variation diminishing (TVD) behavior of the velocity field \citep{2013LeChenadec_Monotonicity,2023Kuhn_MMC}. 
Several studies have attempted to mitigate this issue using dissipative or interface-aware flux limiters \citep{2020Zuzio_CMOM,2023Remmerswaal_Thesis,2023Kuhn_MMC,2021Desmons_HOMP}, but with limited success, largely due to inadequate treatment of interface geometry during velocity reconstruction. 
Only a few works explicitly recognize the need for bounded momentum reconstruction to guarantee velocity boundedness with sharp interfaces \citep{2013LeChenadec_Monotonicity,2017Owkes_UnsplitCMOM,2023Remmerswaal_Thesis}. 
Building on the framework of \citet{2013LeChenadec_Monotonicity}, the present work proposes a generalized \textit{Synchronized Donor Region of Momentum flux} (SynDRoM) method, designed to avoid full interface reconstruction on staggered grids and to integrate naturally within a directionally split framework \citep{2010Weymouth_cVOF}.

\subsection{Treatment of viscosity jumps}
In addition to large density contrasts, high viscosity ratios pose further challenges for the robustness of multiphase flow solvers \citep{2019Nangia_HighDens,2021Desmons_HOMP,2022Li_CMOM}. 
Sharp viscosity jumps induce discontinuities in velocity gradients and complicate the modeling of interfacial viscous stresses, such as those encountered in air lubrication layers \citep{2008Elbing_ALDR}. 
Simple arithmetic averaging of viscosity becomes inadequate in such cases \citep{2011Tryggvason_DNS}.
Alternative approaches, including Ghost Fluid formulations for viscous terms \citep{2015Lalanne_Viscous}, harmonic averaging \citep{2002Prosperetti_Viscosity,2011Tryggvason_DNS,2019Nangia_HighDens}, and anisotropic constitutive laws \citep{2025Magnaudet_ConstitutiveViscous}, have shown limited success, particularly at finite time steps and high Reynolds numbers. 
Other stabilization techniques rely on smoothing material properties, which undermines physical fidelity and conflicts with sharp-interface momentum transport. 
To address this issue, the present study introduces a simple viscosity limiter that constrains the effective kinematic viscosity within a stability-preserving range \citep{2015Campbell_Dissertation}.

\subsection{Structure of the paper}
The paper is organized to progressively identify, analyze, and resolve the key numerical challenges associated with energetic multiphase flows. 
The model is presented as follows:
First, the limitations of velocity-based formulations and the original CMOM scheme are illustrated using a dam-break configuration, supported by energy behaviors.
Second, a simplified one-dimensional test case is then introduced to explain why CMOM in combination with existing interface-aware slope limiters fails to maintain velocity boundedness when sharp interfaces cross the grid boundaries.
Third, building on these observations, we propose a modification of the advection scheme to CMOM, \ie Synchronized Donor Region of Momentum flux (SynDRoM), together with a directionally split momentum update consistent with a conservative Volume-of-Fluid (cVOF) framework and a modified temporal integration strategy.
Finally, the treatment of viscosity jumps is subsequently discussed, leading to the introduction of a viscosity limiter designed to preserve numerical stability without sacrificing physical fidelity.

The validation cases are selected to isolate and examine the physical mechanisms that are relevant to the final demonstration case of standing wave breakers.
The dam-break problem, which also involves strong exchanges between potential and kinetic energy, is first present to compare the characteristic behavior of different numerical schemes.
Highly energetic events also features plentiful fast-traveling droplet; a conceptually related one-dimensional droplet example is employed to clarify the root causes of success or failure for each method.
Since velocity shear is ubiquitous at fluid interfaces and naturally triggers Kelvin-Helmholtz instability, this canonical problem--supported by well-established analytical solutions--is chosen as the validation case for SynDRoM.
Vertical interfaces, which arise during jet formation in wave breaking, motivate the use of a falling-film configuration to assess the robustness of viscosity interpolation schemes.
Finally, the overall capability of the proposed solver is demonstrated through simulations of standing wave breakers, both with and without viscous effects.

\section{Special treatment of advection terms}

\subsection{Limitations of velocity formulations and motivation for a modified CMOM}

\begin{figure*}[ht!]
    \centering
    \begin{subfigure}[b]{0.315\textwidth}
        \centering
        \adjincludegraphics[trim={{.096\width} {.054\height} {.1426\width} {.015\height}}, 
                 clip, height=0.23\textheight]{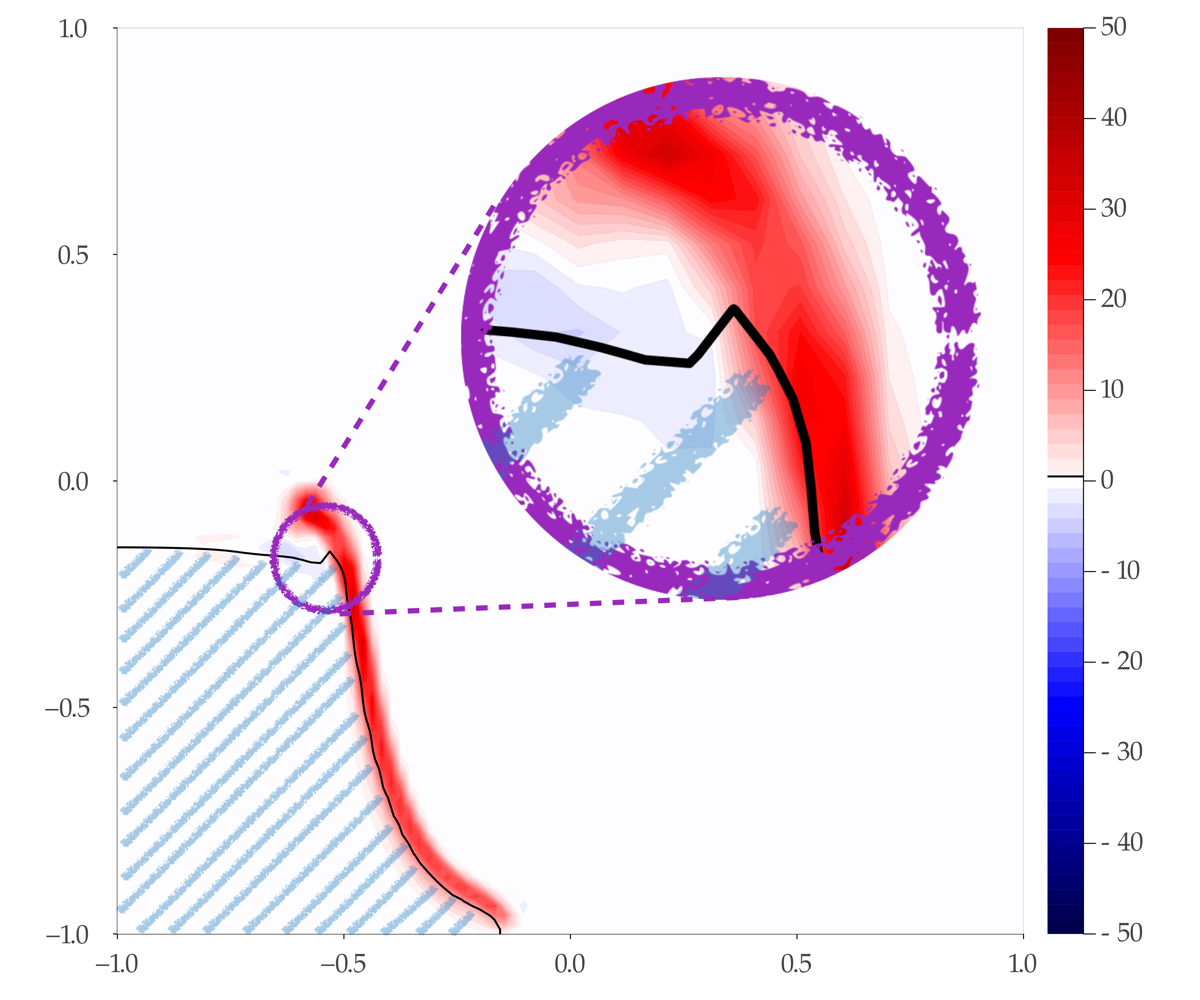}
        \caption{Velocity formulation.}
    \end{subfigure}
    \hfill
    \begin{subfigure}[b]{0.315\textwidth}
        \centering
        \adjincludegraphics[trim={{.096\width} {.054\height} {.1426\width} {.015\height}}, 
                 clip, height=0.23\textheight]{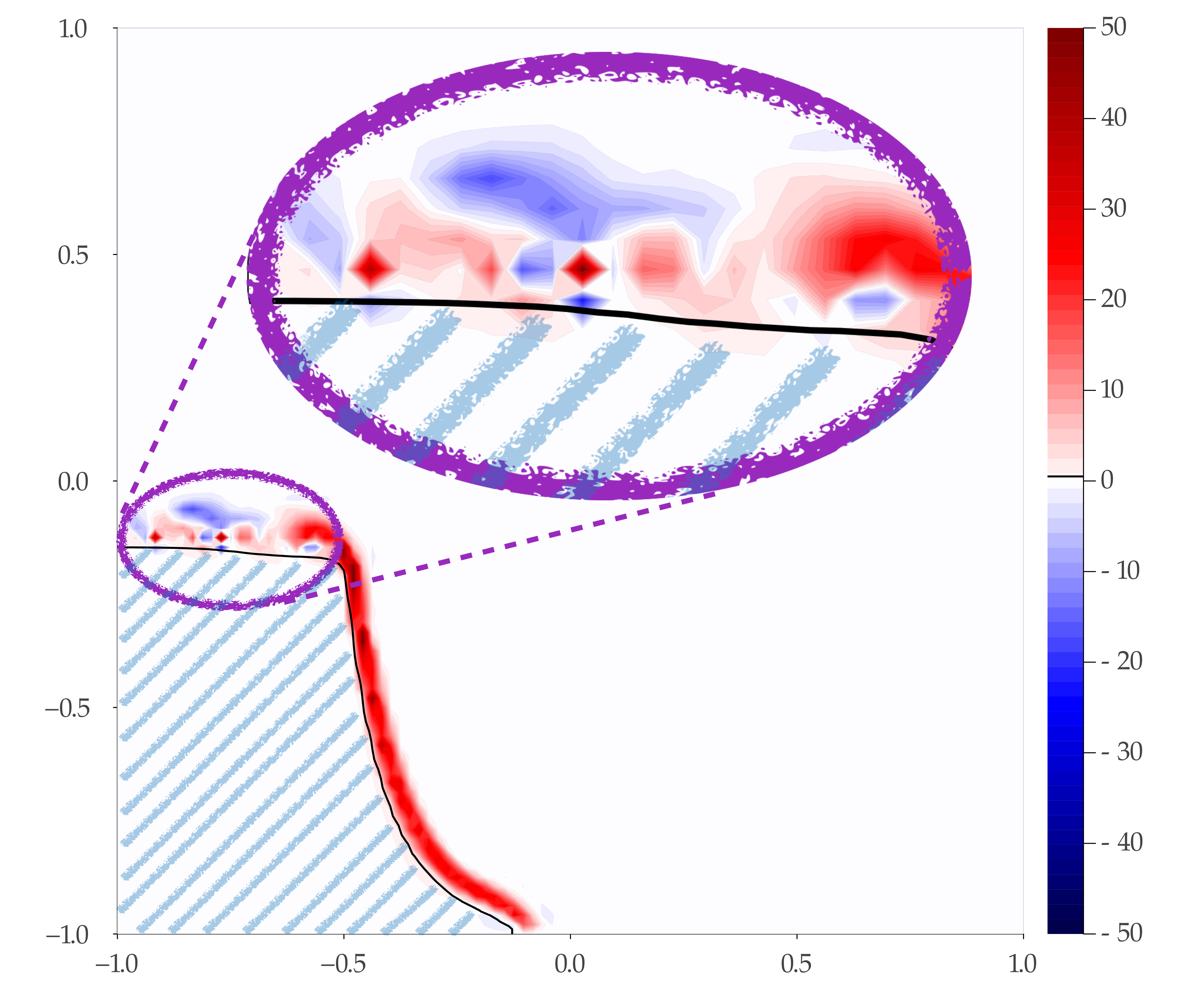}
        \caption{Original CMOM.}
    \end{subfigure}  
    \hfill
    \begin{subfigure}[b]{0.36\textwidth}
        \centering
        \adjincludegraphics[trim={{.096\width} {.054\height} {.0434\width} {.015\height}}, 
                 clip, height=0.23\textheight]{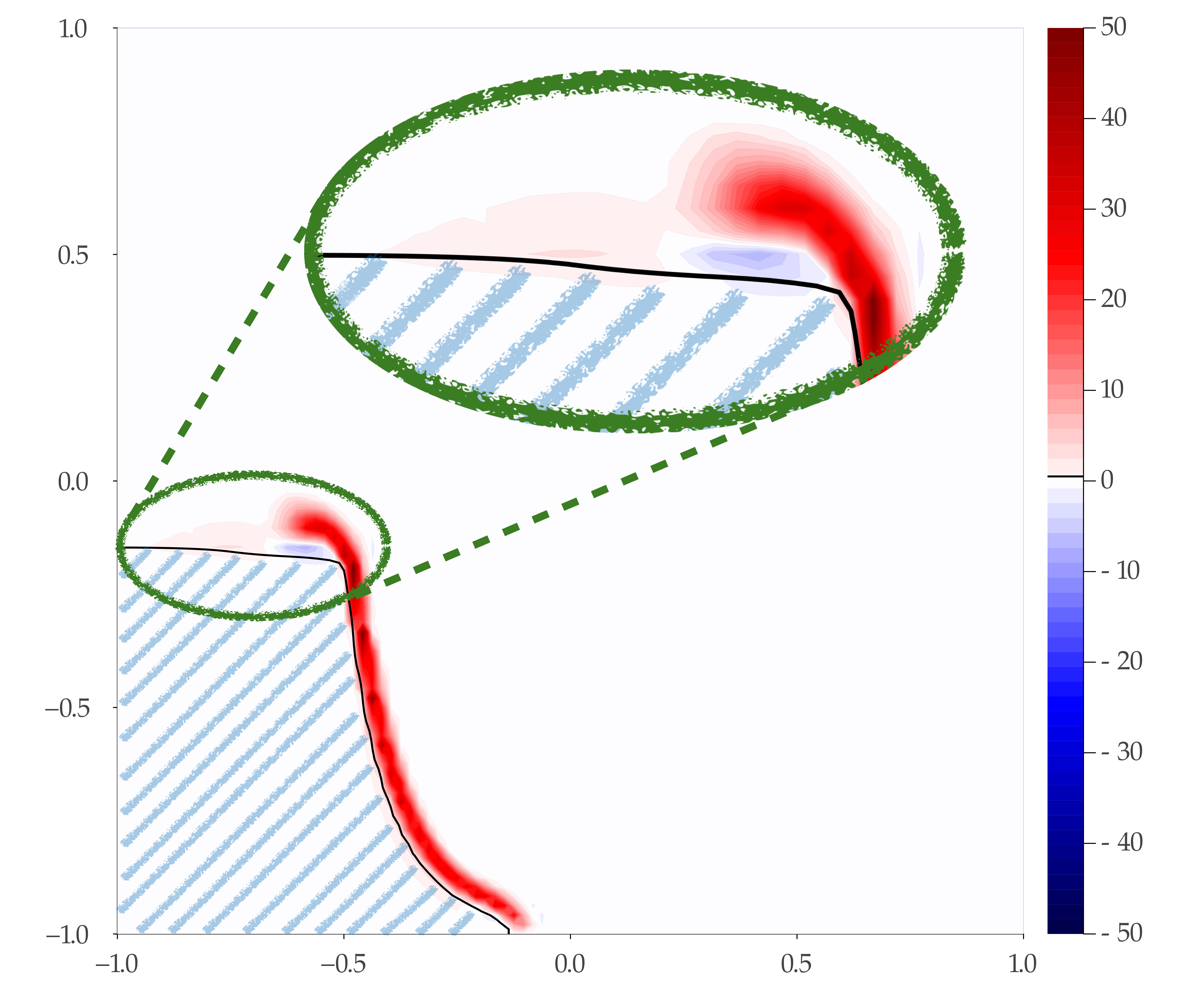}
        \caption{The proposed SynDRoM.}
    \end{subfigure}
    \caption{Interface shape and vorticity field of the 2D dam break example at $t\sqrt{g/H}=0.6$. Teal blue hatches show the water region. Flow field irregularities of velocity and original CMOM formulations indicated and zoomed in with purple oval insets. Flow field of SynDRoM is also enlarged with green oval for comparison purpose. $(N_x,N_y) = (96,96)$.}
    \label{fig:dambreak_shape}
\end{figure*}

\begin{figure}
    \centering
    \includegraphics[width=\linewidth]{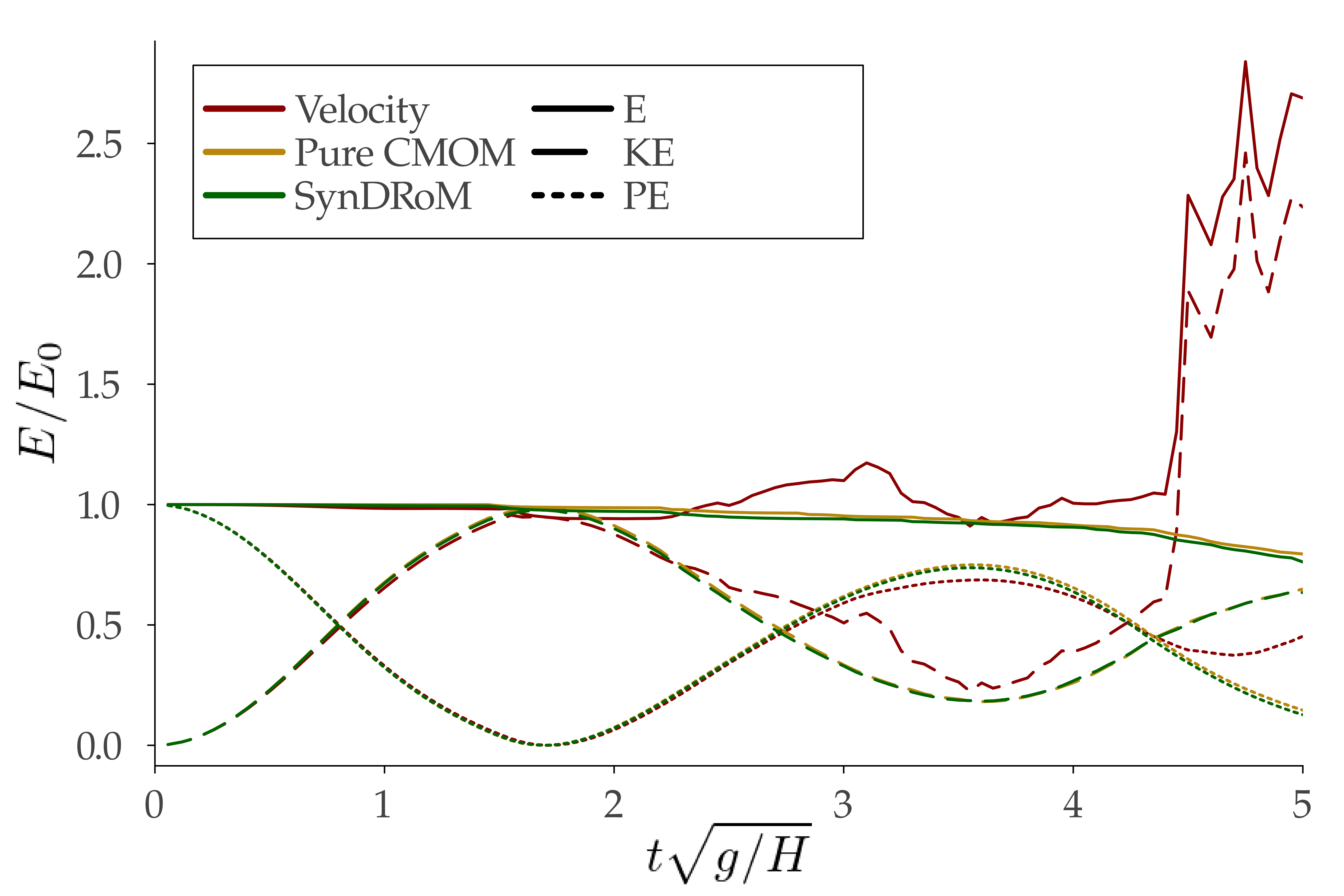}
    \caption{Temporal evolution of energy and its components of the dam break case.}
    \label{fig:dambreak_EnRMS}
\end{figure}

We use a two-dimensional dam-break problem to illustrate the shortcomings of velocity-based formulations and the original CMOM scheme, and to provide an initial comparison with the proposed SynDRoM. 
A column of water of height $H$ collapses under gravity $g$ in the absence of stabilizing mechanisms such as surface tension or viscosity.

\cref{fig:dambreak_shape} compares the interface shape and vorticity field obtained using those three methods, while \cref{fig:dambreak_EnRMS} presents the temporal evolution of mechanical energy and its components. 
In the velocity-based formulation, the air velocity penetrates into the water field, producing an nonphysical devil-horn-shaped interface distortion (highlighted by the purple oval in \cref{fig:dambreak_shape}a). 
This erroneous momentum transfer is accompanied by energy blow-up at later times, as evidenced by the spiky growth of total energy in \cref{fig:dambreak_EnRMS}. 

The original CMOM formulation suppresses velocity penetration and prevents energy divergence; however, it introduces localized velocity blobs in the light phase (marked by purple ovals in \cref{fig:dambreak_shape}b). 
In contrast, the SynDRoM formulation successfully resolves both the interface deformation and the spurious velocity structures without sacrificing the desired physical principles.  
The following sections examine the origin of the velocity blobbing observed in the original CMOM scheme and introduce a remedy.

\subsection{Numerical methods and original CMOM}

\begin{figure}[htb]
    \centering
    \includegraphics[width=\linewidth]{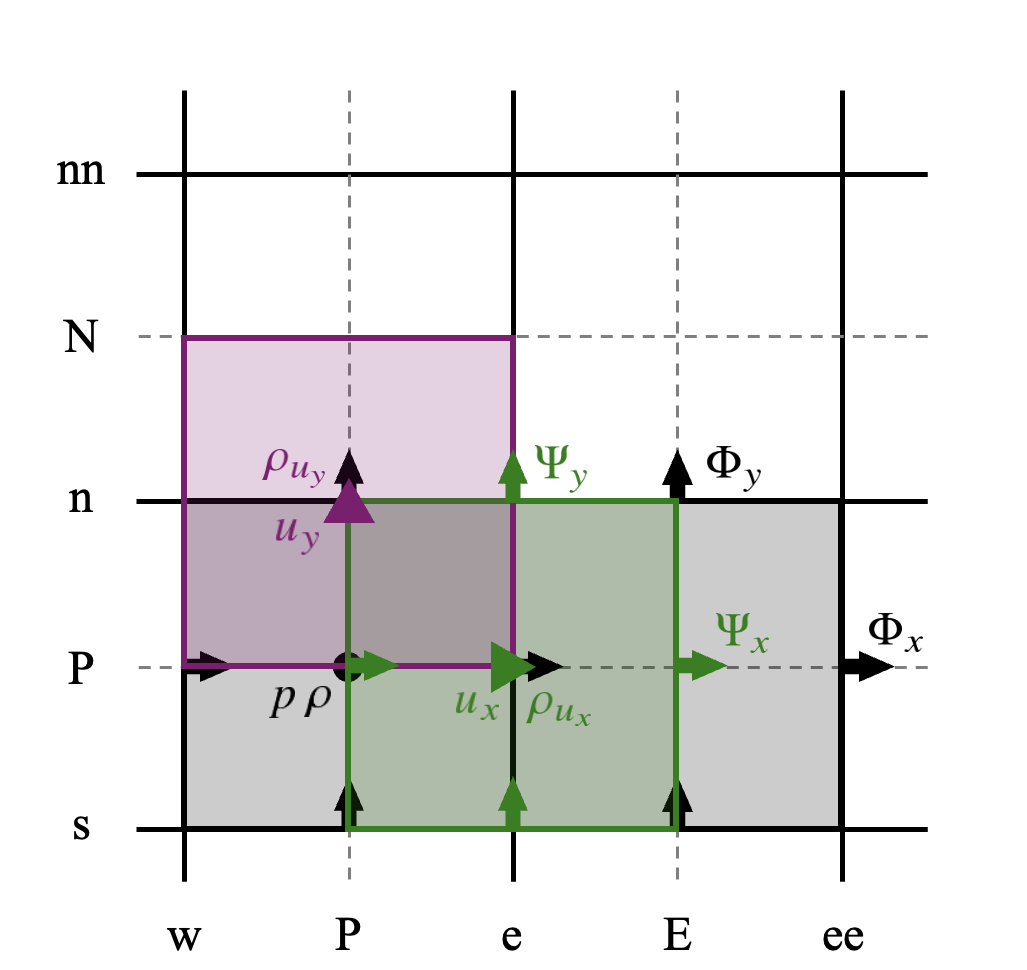}
    \caption{Variable arrangement on a 2D staggered cell complex. The background black cell represents the collocated grid, while the green and purple cells correspond to the horizontal and vertical staggered cells, respectively. We follow Patankar's compass notation for cell and face naming. Pressure $p$ and density $\rho$ are stored in collocated cells, while the velocities $u_i$ and isometric-averaged densities $\rho_{u_i}$ resides on the staggered grids. Mass fluxes on collocated grid's (black, $\Phi_j$) and staggered grid's (green, $\Psi_j$) faces are shown. Only the horizontal momentum control volume (green) is highlighted for clarity, but mass fluxes also reside on faces of the vertical staggered volume (purple).}
    \label{fig:StaggeredGridMassFlux}
\end{figure}

This section summarizes the discretization of advection terms in the original CMOM formulation, providing the necessary background for subsequent analysis. 
Further implementation details are available in the open-source repository \texttt{InterfaceAdvection.jl}\footnote{\url{https://github.com/TzuYaoHuang/InterfaceAdvection.jl}. The code is released under the MIT license and provides simple setup procedures and examples for reproducible simulations and benchmarking. Future developments include the incorporation of Boundary-Data Immersion Methods \citep{2011Weymouth_BDIM} to enable fluid-structure interaction problems such as ship-wave interaction, as well as multi-GPU extensions to support industrial-scale simulations.}, a GPU-accelerated multiphase solver based on \texttt{WaterLily.jl} \citep{2025Weymouth_WaterLily}.

We consider the incompressible one-fluid Navier-Stokes equations, treating the two fluids as a single continuum with spatially varying properties determined by an indicator function. 
The continuity equation is solved using a directionally split conservative Volume-of-Fluid (cVOF) method \citep{2010Weymouth_cVOF}. 
A staggered grid arrangement is adopted (\cref{fig:StaggeredGridMassFlux}) to improve pressure-velocity coupling and preserve favorable spectral properties.

The conservative governing equations required for CMOM are
\begin{alignat}{3}
    &\partial_t \rho &&+ \partial_j \Phi_j &&= 0, \\
    &\partial_t q_i &&+ \partial_j \Psi_j u_i &&= -\partial_i p + \partial_j(\mu(\partial_j u_i + \partial_i u_j)) + g_i,
\end{alignat}
where $q_i=\rho u_i$ denotes momentum. 
The mass fluxes $\Phi_j$ and $\Psi_j$ are defined on the faces of the collocated and staggered grids, respectively. 
$\mu$ is the dynamic viscosity and $g_i$ the body force vectors.
The advected velocity $u_i$ is reconstructed using slope limiters within the implicit large-eddy simulation (iLES) framework. 
Viscous and body-force terms are omitted here and revisited in the viscosity section.

CMOM enforces consistency between the mass fluxes in the continuity and momentum equations by synchronizing $\Phi_j$ and $\Psi_j$, thereby embedding mass conservation into the momentum update.
However, the staggered grid arrangement (\cref{fig:StaggeredGridMassFlux}) prevents direct reuse of collocated mass fluxes in the momentum advection step. 
Several remedies have been proposed, including additional staggered-grid mass advection \citep{2020Zuzio_CMOM}, multi-resolution cell complexes \citep{1998Rudman}, and isometric averaging of collocated-grid fluxes \citep{2003Verstappen_SymmetryPreserving,2021Mirjalili_EConserv,2023Valle_EConserv}. 
In this work, we adopt the latter approach due to its simplicity and strict mass and momentum conservation.

Isometric averaging--\ie: volumetric averaging of density and area-weighted averaging of mass flux--maps collocated quantities onto staggered control volumes while preserving mass conservation by construction. 
For the horizontal staggered cell in \cref{fig:StaggeredGridMassFlux}, the corresponding quantities are
\begin{align*}
    \rho_{u_x;\mathrm{e}} &= (
    \rho_{\mathrm{P}} \mathcal{V}_{\mathrm{P}}  + \rho_{\mathrm{E}} \mathcal{V}_{\mathrm{E}}
    )/(
    \mathcal{V}_{\mathrm{P}}+\mathcal{V}_{\mathrm{E}}
    ),\\
    \Psi_{x;\mathrm{E}} &= (
    \Phi_{x;\mathrm{e}} \mathcal{A}_{\mathrm{e}} + \Phi_{x;\mathrm{ee}} \mathcal{A}_{\mathrm{ee}}
    )/(
    \mathcal{A}_{\mathrm{P}}+\mathcal{A}_{\mathrm{ee}}
    ), \\
    \Psi_{y;\mathrm{ne}} &= (
    \Phi_{y;\mathrm{n}} \mathcal{A}_{\mathrm{n}} + \Phi_{y;\mathrm{nE}} \mathcal{A}_{\mathrm{nE}}
    )/(
    \mathcal{A}_{\mathrm{n}}+\mathcal{A}_{\mathrm{E}}
    ),
\end{align*}
where $\mathcal{A}$ and $\mathcal{V}$ denote face areas and cell volumes. 
For simplicity, a uniform unit cell complex is assumed hereinafter; extensions to stretched grids and three dimensions are straightforward.

\subsection{Failure of original CMOM: non-monotonic transport} \label{sec:example}

\begin{figure*}
    \centering
    \includegraphics[width=0.7\textwidth]{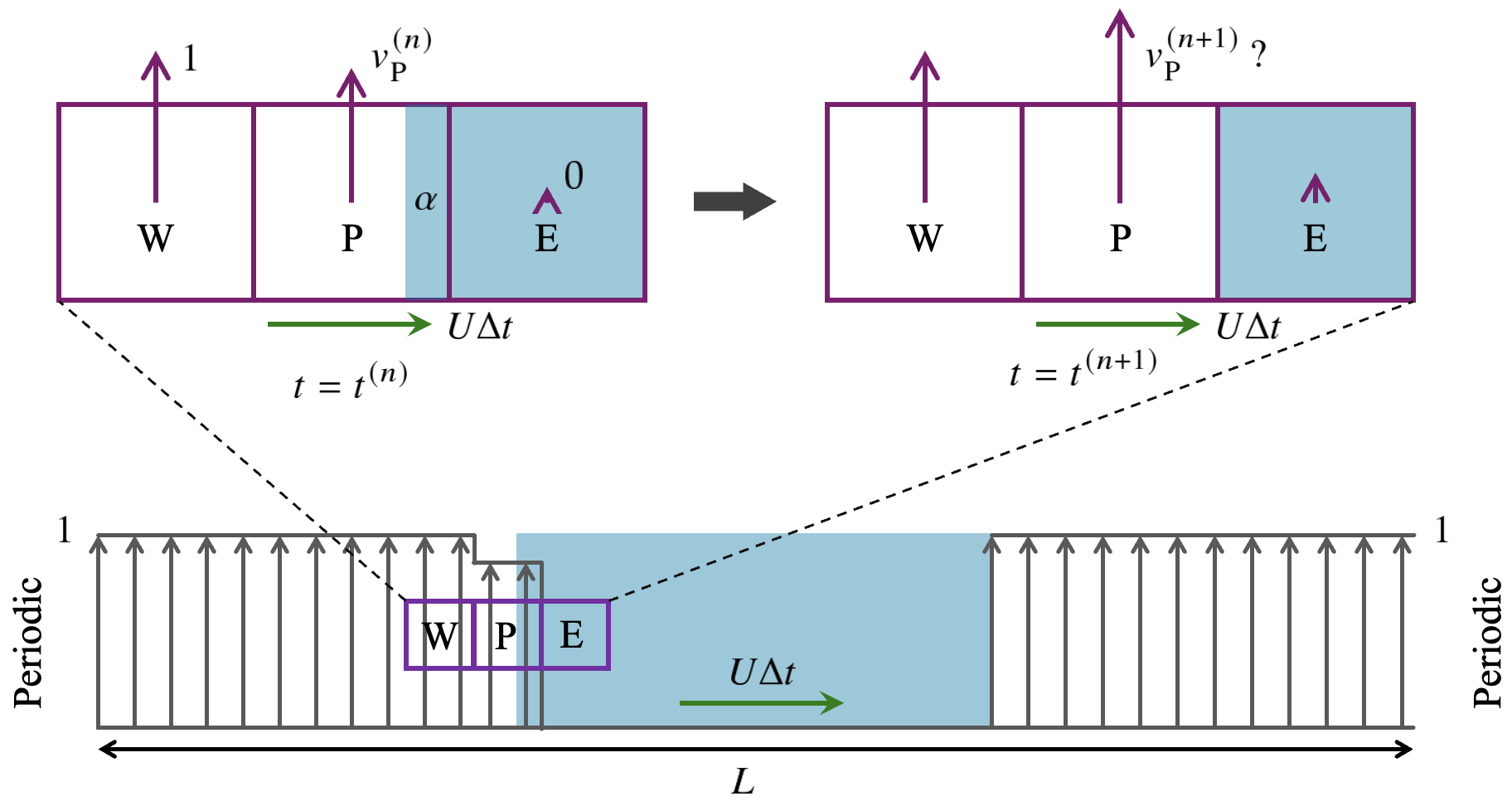}
    \caption{Quasi-one-dimensional high density droplet (teal blue shade) transport. The high vertical shear present at the rear interface of the droplet (grid cell $\mathrm{P}$) and is transported by $U$. After one time step $\Delta t$, the horizontal velocity transfer all but no more than the remaining high density fluid in $\mathrm{P}$ to $\mathrm{E}$ -- the volume fraction inside the central satisfies $\alpha = U\Delta t/\Delta x$. The new vertical velocity of the central cell $\mathrm{P}$ is to be determined. The velocity omits tensor notation and use $v$ to indicate the vertical one.}
    \label{fig:AliasingCase}
\end{figure*}

\begin{figure*}
    % \begin{subfigure}[t]{0.47\textwidth}
    %     \centering
    %     \includegraphics[width=0.9\textwidth]{gfx/Vel-CMOM-All_initial.png}
    %     \caption{Initial condition of vertical velocity, momentum and volume fraction.}
    % \end{subfigure}
    % \hfill
    \begin{subfigure}[t]{0.47\textwidth}
        \centering
        \includegraphics[width=\textwidth]{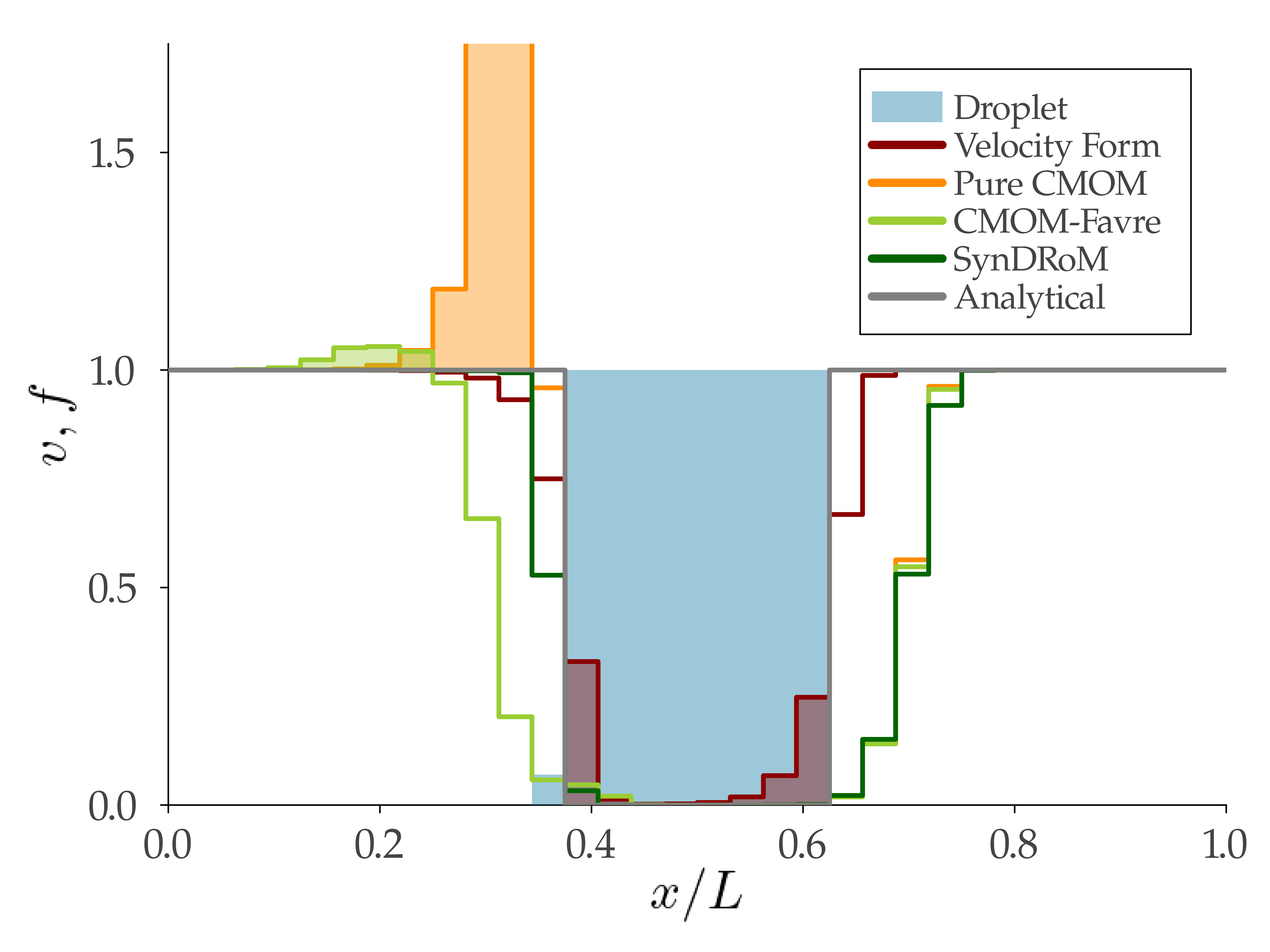}
        \caption{Velocity profile after one convection cycle.}
        \label{fig:DropletIaFL_end}
    \end{subfigure}
    \begin{subfigure}[t]{0.47\textwidth}
        \centering
        \includegraphics[width=\textwidth]{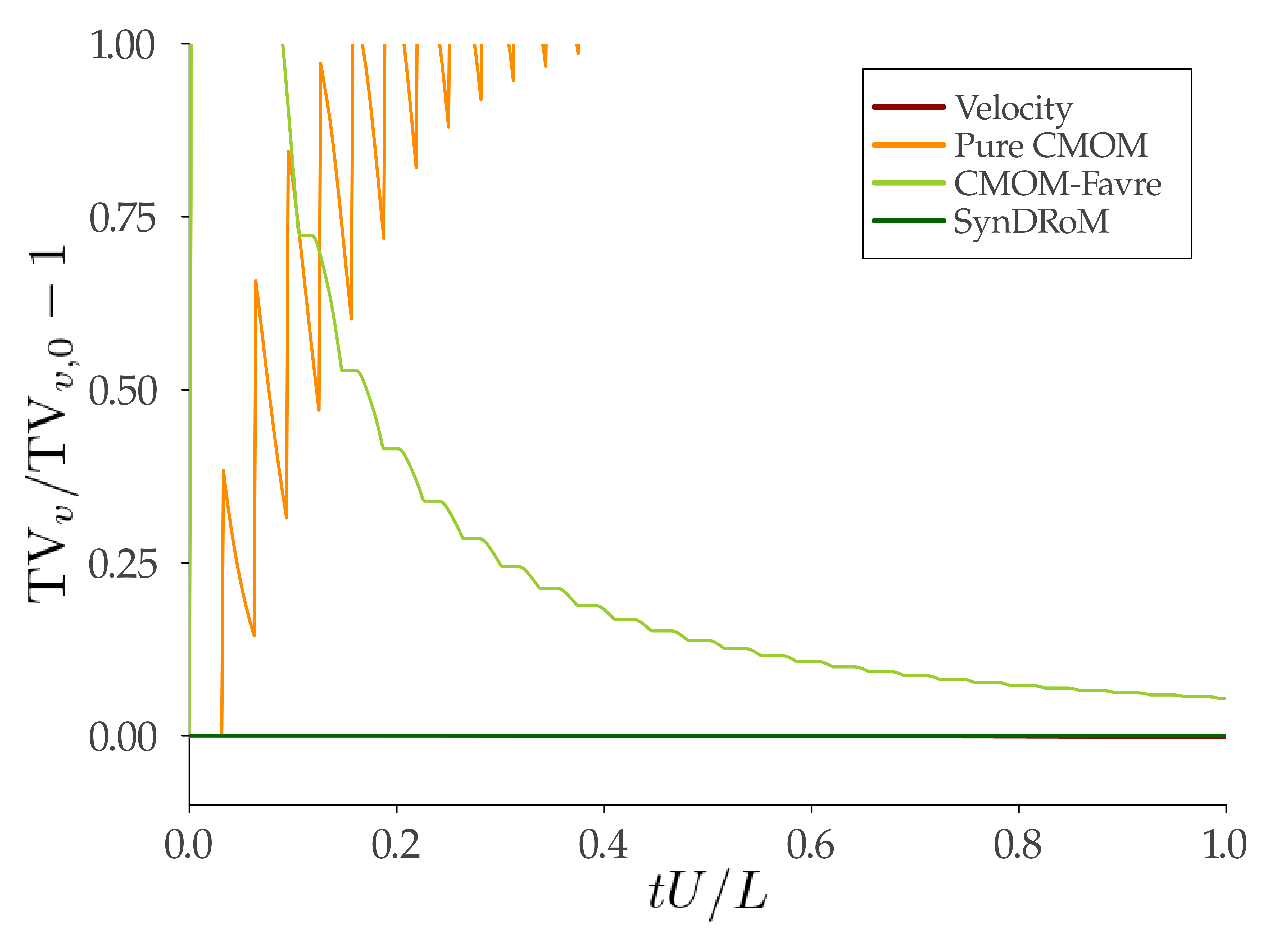}
        \caption{Total variation of vertical velocity.}
    \end{subfigure}
    \caption{One-dimensional periodic droplet advection of density ratio $\lambda_\rho=10^{-3}$. Initial vertical velocity (gray line) has a (inverse) top-hat profile with a little overlap with the droplet (teal blue shade); the field is convected using forward Euler and superbee interpolation. Result and comparison are present after one convection period ($tU/L = 1$). $N_x=32$, $\mathrm{CFL}=U\Delta t/\Delta x = 0.05$. The shades of CMOM variants show the overshoots while those of velocity formulation show the velocity leaks.}
    \label{fig:DropletIaFL}
\end{figure*}

The quasi-one-dimensional test case shown in \cref{fig:AliasingCase}, adapted from \citet{2023Kuhn_MMC,2021Arrufat_MassMomentumConserv}, illustrates the failure of original CMOM schemes--both with traditional and existing interface-aware slope limiters--to preserve velocity monotonicity. 
The setup is as follows: a dense, 1D, droplet embedded in air is advected uniformly to the right with velocity $U$, while a sharp vertical velocity gradient exists at the rear interface. 
This configuration represents an extreme but revealing edge case, characterized by a localized spike in momentum.
The flow behavior is conceptually similar to the top interface of dam break after a few time steps, with velocity pointing downwards.

The simplified governing equations reduce to
\begin{alignat*}{3}
    \partial_t &\rho + &&\partial_j \Psi &&= 0, \\
    \partial_t &q + &&\partial_j \Psi v &&= 0,
\end{alignat*}
where $\Psi=\rho U$ is the horizontal mass flux and $q=\rho v$ is the vertical momentum. Simulation results are summarized in \cref{fig:DropletIaFL}. 
In the inviscid limit, the vertical velocity should be transported exactly, maintaining a bounded top-hat profile without overshoot.

Velocity-based schemes, although effective for scalar transport, allow spurious momentum leakage from the light to the dense phase, leading to incorrect momentum and energy transfer and ultimately distorting the interface (\textit{cf.} \cref{fig:dambreak_shape}a). 
The original CMOM formulation and its Favre-averaged variants \citep{2023Kuhn_MMC} eliminate velocity penetration but instead generate nonphysical velocity overshoots in the light phase, violating the TVD condition. 
Favre averaging additionally introduces excessive diffusion at later stages. 
In contrast, the proposed method maintains monotonicity through controlled upwind diffusion localized at the interface.

To elucidate the failure mechanism, we examine the three cells adjacent to the rear of the momentum spike (\cref{fig:AliasingCase}). 
After one time step, the dense phase in cell $\mathrm{P}$ is fully advected out, corresponding to $\alpha=U\Delta t/\Delta x$. 
With $v_{\mathrm{W}}^{(n)}=1$, $v_{\mathrm{E}}^{(n)}=0$, $v_{\mathrm{E}}^{(n)} \le v_{\mathrm{P}}^{(n)} \le v_{\mathrm{W}}^{(n)}$, and density ratio $\lambda_\rho\ll1$, the physical-relevant solution requires
\[
v_{\mathrm{E}}^{(n)} \le v_{\mathrm{P}}^{(n+1)} \le v_{\mathrm{W}}^{(n)} .
\]
However, in CMOM the updated velocity is obtained by dividing the advected momentum by the updated density, which leads to
\begin{align*}
v_{\mathrm{P}}^{(n+1)}
= \underbrace{\alpha (1-v_{\mathrm{P}}^{(n)}) + v_{\mathrm{P}}^{(n)}}_\text{bounded}
+ \underbrace{(1-\gamma)v_{\mathrm{P}}^{(n)}\alpha/\lambda_\rho}_\text{unbounded}.
\end{align*}
$\gamma$ is introduced to represent different choice of interpolation at east face such as upwind ($\gamma = 1$), central ($\gamma = 1/2$) and downwind ($\gamma = 0$). 
Note that we omit the interpolation at face w for simplicity.
For small density ratio, $\lambda_\rho$, even modest deviations from upwind interpolation ($1-\gamma\gg\lambda_\rho$) result in large velocity overshoots. 
Favre-averaged slope limiting \citep{2023Kuhn_MMC} reduces but does not eliminate this effect, as the effective interpolation coefficient $\gamma$ remains too large to satisfy the TVD condition.

The underlying issue is the lack of synchronicity between the advected momentum and the mass flux responsible for its transport, leaving excessive momentum in cells containing vanishingly small mass, leading to a spike in velocity. 
While this error diminishes for extremely small time steps ($U\Delta t/\Delta x=\mathcal{O}(\lambda_\rho)$), such restrictions are impractical. 
To retain reasonable time-step sizes while enforcing velocity monotonicity, we introduce the synchronized momentum donor method in the following section.

\subsection{Synchronized Donor Region of Momentum Flux (SynDRoM)} \label{sec:CoMaSL}

\begin{figure*}
    \centering
    \includegraphics[width=0.65\linewidth]{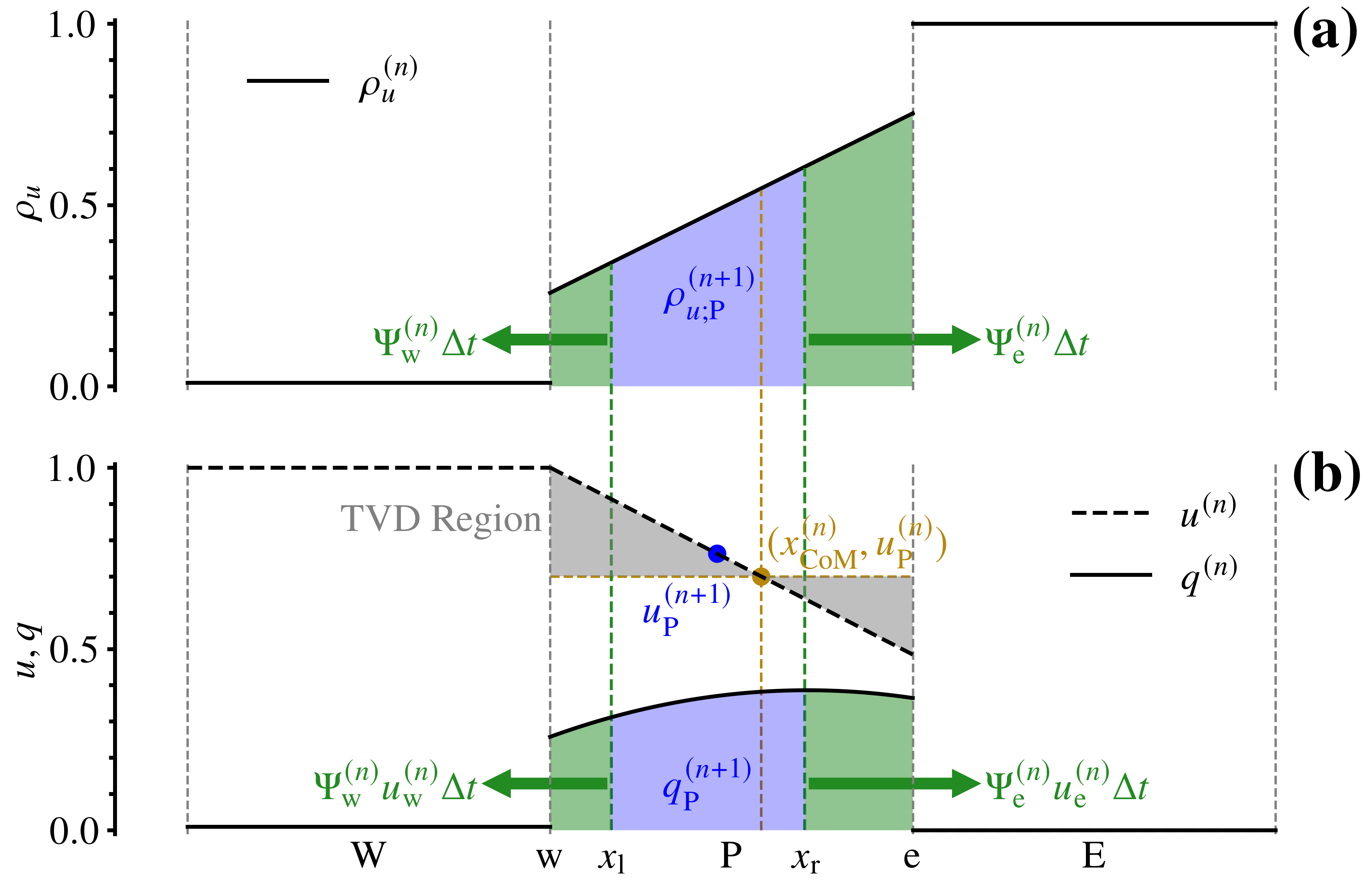}
    \caption{Mass-flux-based donor region concept in SynDRoM with linear density distribution assumption, example using horizontal mass flux with focus on cell $\mathrm{P}$.
    Panel (a) shows the density distribution and mass flux and (b) those for velocity and momentum.
    % The slopes of density and velocity in cell $\mathrm{P}$ are estimated with superbee slope limiter concept.
    Center of mass are labeled with vertical mustard line, and the level of original velocity is signified with horizontal mustard line in (b). 
    Lines pivoting around the mustard point are allowed in TVD conditions, the black line on the boundary is an example of using superbee philosophy.
    The green region in (a) represent the outing mass fluxes ($\Psi^{(n)}_\mathrm{w/e} \Delta t$), while the blue represents the updated mass, $\rho_{u;\mathrm{P}}^{(n+1)}$. 
    The mass flux donor region is synchronous to the momentum distribution to identify the correct momentum flux region: shown with two green dashed lines, $x_\mathrm{l}$ and $x_\mathrm{r}$ spanning across (a) and (b).
    The updated velocity $u^{(n+1)}_{\mathrm{P}}$, are weighted averaged of the rest velocity.
     }
    \label{fig:CoMaSL_SlopeDensity}
\end{figure*}

\begin{figure*}
    \centering
    \includegraphics[width=0.65\linewidth]{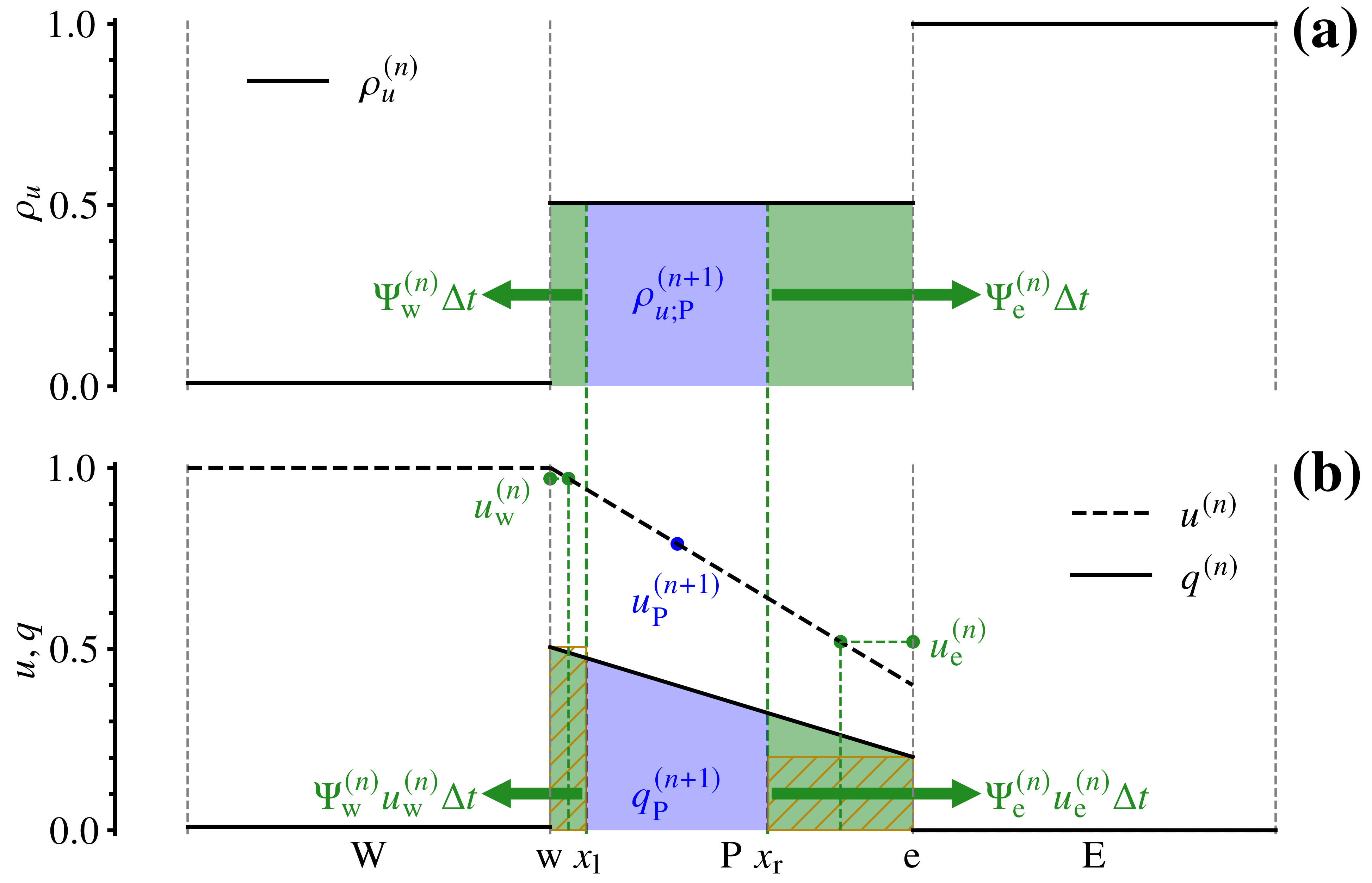}
    \caption{All is similar to \cref{fig:CoMaSL_SlopeDensity} but the uniform density assumption.
    The mass flux donor region is extended to the momentum distribution to identify the correct momentum donor region (two green trapezoids).
    The effective advected velocities, $u^{(n)}_{\mathrm{w/e}}$, are midpoint of the contained velocity thanks to uniform density assumption.
    The orange hatch shows the advection sweep of the original CMOM with the given slope limiter, where it over-advect on west face and under-advect on the east face.
     }
    \label{fig:CoMaSL_UniformDensity}
\end{figure*}

To achieve fully synchronized momentum transport, density, velocity, and momentum must be reconstructed consistently within each cell \citep{2013LeChenadec_Monotonicity,2017Owkes_UnsplitCMOM}. 
The proposed Synchronized Donor Region of Momentum flux (SynDRoM) follows the spirit of LeVeque’s Reconstruct-Evolve-Average (REA) framework \citep{2002LeVeque}. 
Momentum is first reconstructed within a cell, evolved using a mass-flux-based donor concept, and finally averaged to obtain an updated velocity that satisfies the TVD condition. 
In contrast to many existing interface-aware slope limiters, which focus solely on velocity reconstruction, SynDRoM explicitly accounts for the momentum distribution and the corresponding donor region, determined by mass transport.

We illustrate the method using a momentum control volume $q_{\mathrm{P}}^{(n)}$ undergoing horizontal advection, as shown in \cref{fig:CoMaSL_SlopeDensity}. 
Tensor notation is omitted for clarity. 
SynDRoM consists of two steps:
\begin{itemize}
    \item \textit{Momentum reconstruction.}  
    A density-velocity distribution pair $(\rho(x),u(x))$ is reconstructed such that it reproduces the cell-averaged density and momentum,
    \begin{align*}
        \rho_{u;\mathrm{P}}^{(n)} &= \int_{x_\mathrm{w}}^{x_\mathrm{e}} \rho(x)\,\dd{x}, \\
        q^{(n)}_\mathrm{P} &= \int_{x_\mathrm{w}}^{x_\mathrm{e}} \rho(x)u(x)\,\dd{x},
    \end{align*}
    while remaining bounded by neighboring cell values in the advection direction to satisfy the TVD constraint.
    Since these conditions admit infinitely many solutions, a practical strategy is adopted: a density distribution (being uniform, sloped, or Heaviside, \textit{etc.}) is first assumed, from which the center of mass $x_\mathrm{CoM}$ is determined (vertical mustard line in \cref{fig:CoMaSL_SlopeDensity}). 
    A linear velocity distribution is then constructed by pivoting about $(x_\mathrm{CoM},u_{\mathrm{P}}^{(n)})$ (mustard point in panel \cref{fig:CoMaSL_SlopeDensity}b), constrained by neighboring bounds. 
    Any such reconstruction within the admissible envelope (gray area in \cref{fig:CoMaSL_SlopeDensity}b) yields a TVD update.

    \item \textit{Momentum evolution with a synchronized donor region.}  
    The mass flux $\Psi^{(n)}$ is already known from the preceding mass-advection step and has been isometrically averaged onto the staggered grid. 
    To ensure synchronicity, the donor region for momentum must correspond exactly to the mass-flux donor region. 
    For the east face $\Psi_{\mathrm{e}}^{(n)}$, the right boundary $x_\mathrm{r}$ of the donor region is determined such that
    \begin{align*}
        \Psi_{\mathrm{e}}^{(n)}\Delta t = \int_{x_\mathrm{r}}^{x_\mathrm{e}} \rho(x)\,\dd{x}.
    \end{align*}
    The momentum flux is then evaluated over the very same region,
    \begin{align}
        \Psi_{\mathrm{e}}^{(n)}u_{\mathrm{e}}^{(n)}\Delta t
        = \int_{x_\mathrm{r}}^{x_\mathrm{e}} \rho(x)u(x)\,\dd{x},
    \end{align}
    yielding the effective advected velocity being in the form of weighted average
    \begin{align}
        u_{\mathrm{e}}^{(n)}
        = \left.\int_{x_\mathrm{r}}^{x_\mathrm{e}} \rho(x)u(x)\,\dd{x}\right/
        \int_{x_\mathrm{r}}^{x_\mathrm{e}} \rho(x)\,\dd{x}. \label{eq:EffectiveUe}
    \end{align}
    After update, the remaining velocity, also the weighted-average of the bounded velocity distribution ($u(x)$), is obviously bounded due to the properties of convex combination.
\end{itemize}
The same procedure also applies to diverging mass fluxes, as illustrated in \cref{fig:CoMaSL_SlopeDensity}, while still preserving velocity boundedness through synchronized reconstruction and transport.

Interestingly, the advected velocity in \cref{eq:EffectiveUe} will approach upwind solution if the outing mass approach the original mass content, \ie $x_r \rightarrow x_w$.
This explains why existing interface-aware flux limiters tend to apply dissipative flux limiters near the interfaces.

The proposed approach, unlike \citet{2013LeChenadec_Monotonicity,2017Owkes_UnsplitCMOM}, allows the density distribution used for momentum transport to differ from the geometric interface representation, enabling compatibility with isometric interpolation and staggered grids.
The donor region is mass-flux-based rather than volume-based, and thus does not need to coincide with the conventional advection sweep near the interface.

For simplicity, we further assume a uniform density distribution within staggered momentum cells (\cref{fig:CoMaSL_UniformDensity}). 
This assumption is reasonable, as mapping density from collocated to staggered grids inevitably loses some geometric detail of sharp reconstructions. 
With the center of mass located at the cell center, standard single-phase slope limiters can be applied to reconstruct the velocity. 
The additional cost due to SynDRoM's modification on slope limiter is thus small compared to other major computational cost such as the reconstruction step in geometric Volume-of-Fluid methods and the pressure projection step.
The assumption also enables us to compare with the original CMOM.
The orange hatched regions in \cref{fig:CoMaSL_UniformDensity} illustrate how the original CMOM scheme over- and under-advects momentum when the donor-region concept is not enforced, leading to oscillatory velocity fields.

\subsection{Directionally split momentum update: consistency with mass transport} \label{sec:dirsplit}

\begin{figure*}[tb]
    % First row: (a) and (b)
    \begin{subfigure}[b]{0.49\textwidth}
        \centering
        \includegraphics[width=0.8\textwidth]{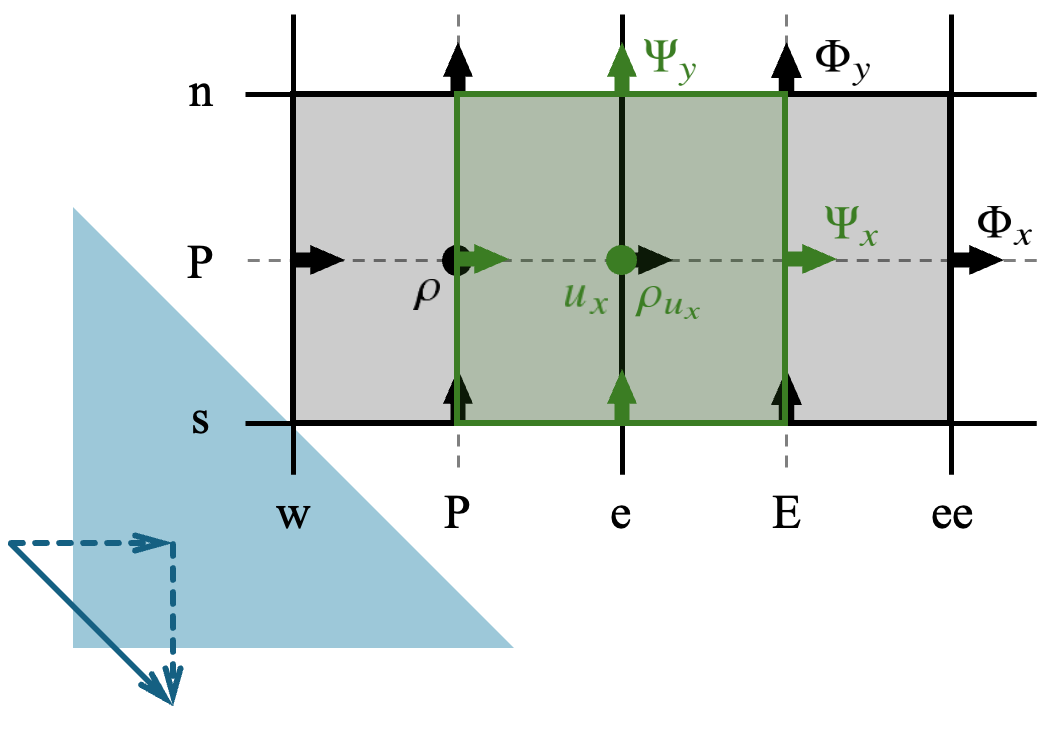}
        \caption{$t^{(n)}$}
        \label{fig:DirSplitCase-a}
    \end{subfigure}
    \hfill
    \begin{subfigure}[b]{0.49\textwidth}
        \centering
        \includegraphics[width=0.8\textwidth]{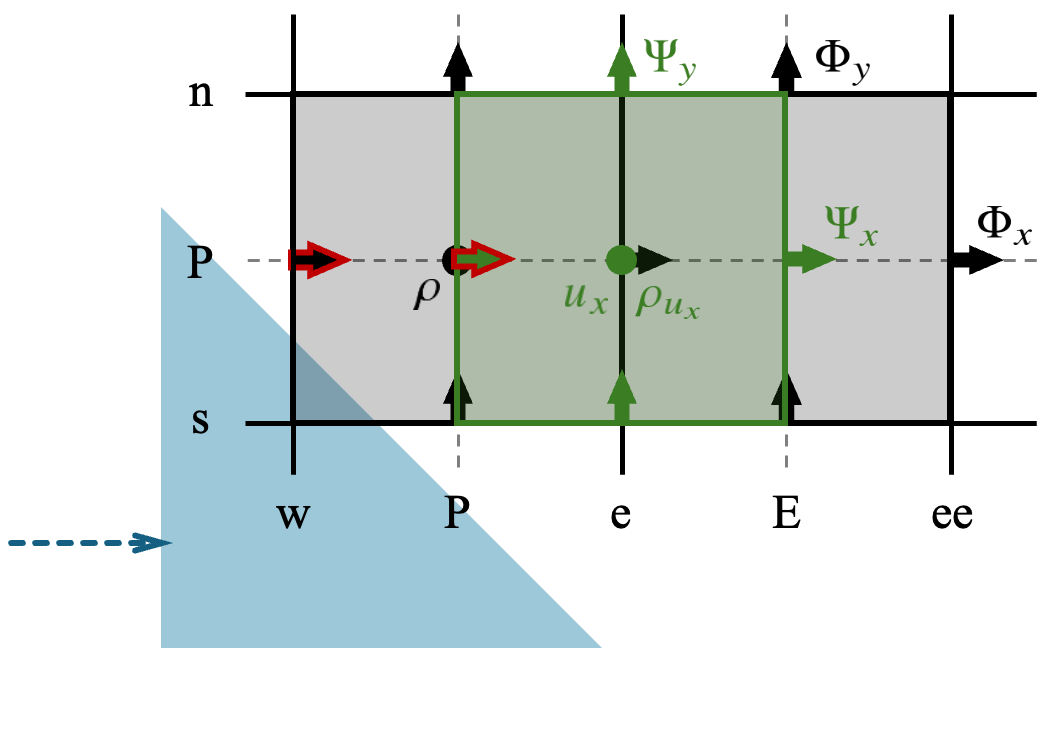}
        \caption{$t^{(n+1,x)}$}
        \label{fig:DirSplitCase-b}
    \end{subfigure}
    \begin{subfigure}[c]{0.49\textwidth}
        \centering
        \includegraphics[width=0.5\textwidth]{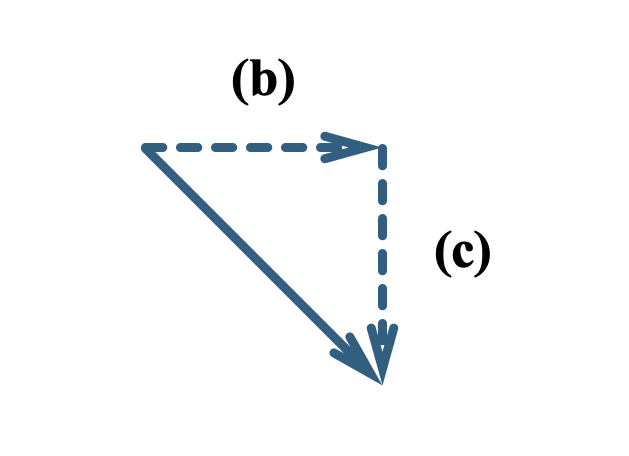}
    \end{subfigure}
    % Row 2: (c), aligned right
    \hfill
    \begin{subfigure}[c]{0.49\textwidth}
        \centering
        \includegraphics[width=0.8\textwidth]{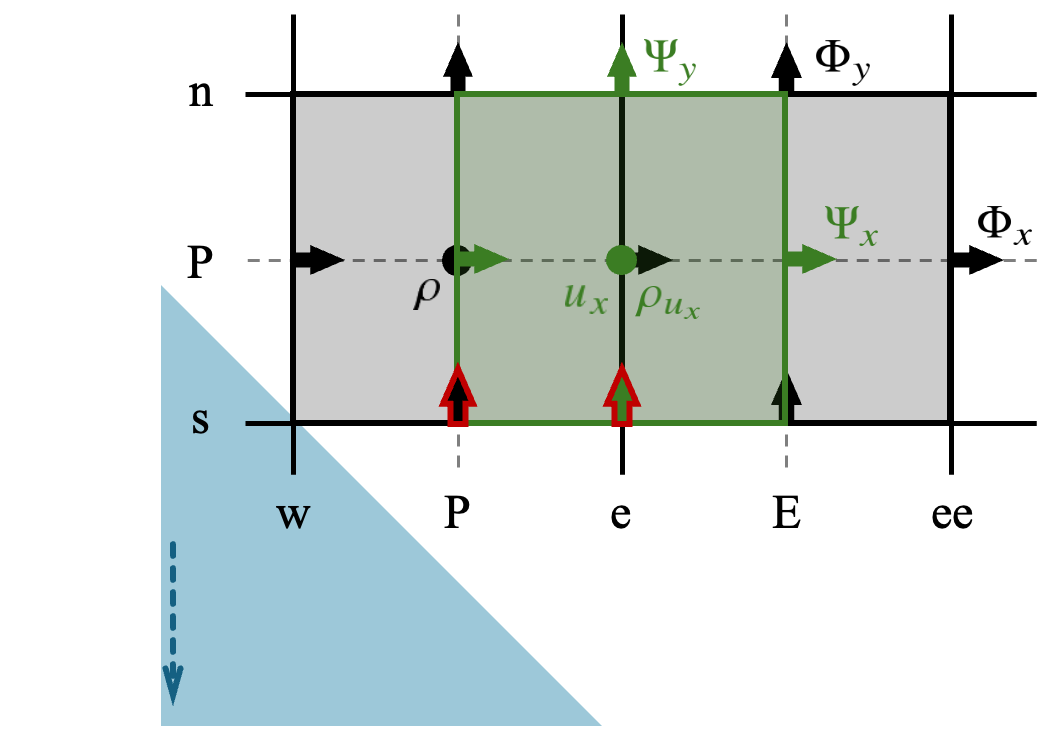}
        \caption{$t^{(n+1)} = t^{(n+1,y)}$}
        \label{fig:DirSplitCase-c}
    \end{subfigure}
    \caption{Directionally-split mass update of a droplet's side tangentially passing through a collocated cell. The figures are presented in visual order. The solid line represents the true motion path while the dash lines signify the auxiliary path from directional split algorithm. The $x$ sweep (b) is performed first and then the $y$ one (c). Bold red outlines around the arrows show the influenced mass fluxes.}
    \label{fig:DirSplitCase}
\end{figure*}

As noted by \citet{2021Arrufat_MassMomentumConserv}, a directionally split momentum update is required to remain consistent with the directionally split mass advection in cVOF. 
Attempts to avoid this through transverse-gradient corrections \citep{2023Kuhn_MMC} have shown limited effectiveness.

Consider the example in \cref{fig:DirSplitCase}, where a dense fluid mass passes tangentially through an initially empty cell over one time step. 
Because the mass advection proceeds sequentially in $x$ and $y$, large intermediate mass fluxes appear on certain staggered faces (highlighted by red arrows). 
If momentum were updated in an unsplit fashion while using those mass fluxes, these fictitious fluxes could exceed the original available mass in the staggered control volume, violating velocity boundedness. 
In essence, momentum would be transported where no mass yet exists.

To avoid this inconsistency, momentum transport is further synchronized with each directional mass-advection substep, including the dilation correction inherent to cVOF. 
A directional update in the $j$-th direction reads
\begin{align} 
    \rho^{(n+1,j)} &= \rho^{(n+1,j^\prime)} + \Delta t\partial_j^h \Phi_j^{(n+1,j)} \nonumber\\
    &\phantom{=}+ \Delta t \, \rho^{(n)}_c  (\partial_j^h u_j^{(n)}),  \\
    q_i^{(n+1,j)} &= q_i^{(n+1,j^\prime)} + \Delta t\partial_j^h \Psi_j^{(n+1,j)} u_i^{(n+1,j^\prime)} \nonumber \\
    &\phantom{=}+ \Delta t u_i^{(n)}\, \overline{\rho^{(n)}_c  (\partial_j^h u_j^{(n)})},  \\
    u_i^{(n+1,j)} &= q_i^{(n+1,j)} / \rho_{u_i}^{(n+1,j)} ,
\end{align}
where no Einstein summation is implied over $j$. 
Here, $\partial_j^h$ denotes the discrete derivative in direction $j$, and the overline indicates isometric averaging over the collocated cells overlapping the staggered momentum control volume. 
As in directionally split cVOF, the advected velocity is taken from the previous substep $(n+1,j^\prime)$. 
The formulation extends naturally to higher-order operator splitting schemes, such as Strang splitting \citep{1968Strang_StrangSplit} and Auzinger-Ketcheson splitting \citep{2017Auzinger_Split}, with appropriate ordering and time-step control. 
In this work, a permuted Strang splitting is employed.

\subsection{Temporal integration: density-weighted RK2}

To preserve interface sharpness, a second-order Runge-Kutta method is adopted instead of Heun’s predictor-corrector scheme. 
In multiphase flows with CMOM, however, the midpoint velocity cannot be obtained by simple arithmetic averaging. 
Since density and momentum evolve linearly in time but velocity does not, a density-weighted average is required:
\begin{align}
    u^{(n+\nfot)}_i =
    \frac{q^{(n)}_i + q^{(n+1^\circ)}_i}
         {\rho^{(n)}_{u_i} + \rho^{(n+1^\circ)}_{u_i}}
    \neq
    \frac{u^{(n)}_i + u^{(n+1^\circ)}_i}{2}.
\end{align}

\subsection{Two-dimensional Kelvin-Helmholtz instability: preservation of flow physics}

\begin{figure*}[ht!]
    \centering
    \begin{subfigure}[b]{0.18\textwidth}
        \centering
        \adjincludegraphics[trim={{.18375\width} {.055\height} {.24\width} {.015\height}}, 
                 clip, height=0.28\textheight]{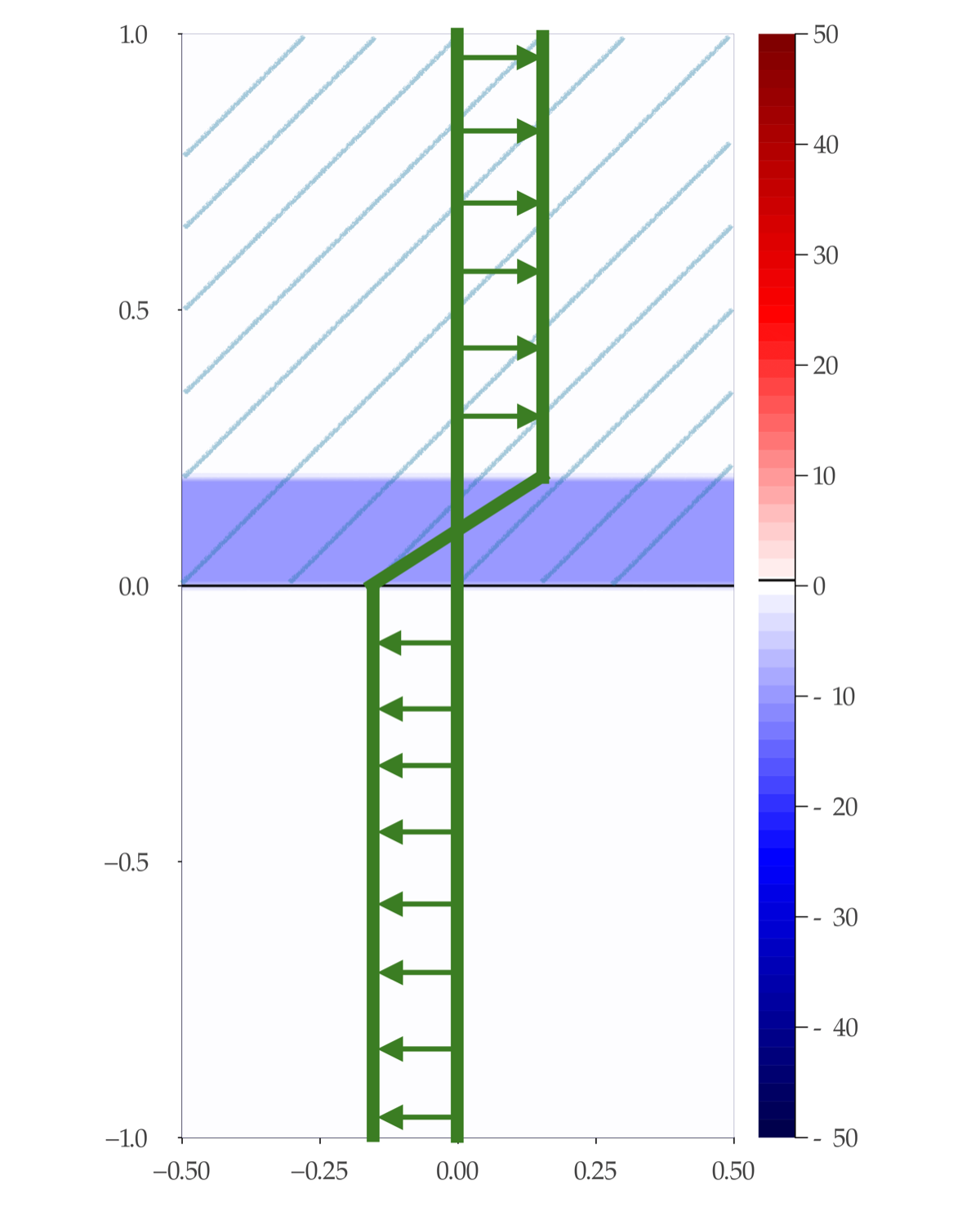}
        \caption{$tU/L=0.0$.}
    \end{subfigure}
    \hfill
    \begin{subfigure}[b]{0.18\textwidth}
        \centering
        \adjincludegraphics[trim={{.18375\width} {.055\height} {.24\width} {.015\height}}, 
                 clip, height=0.28\textheight]{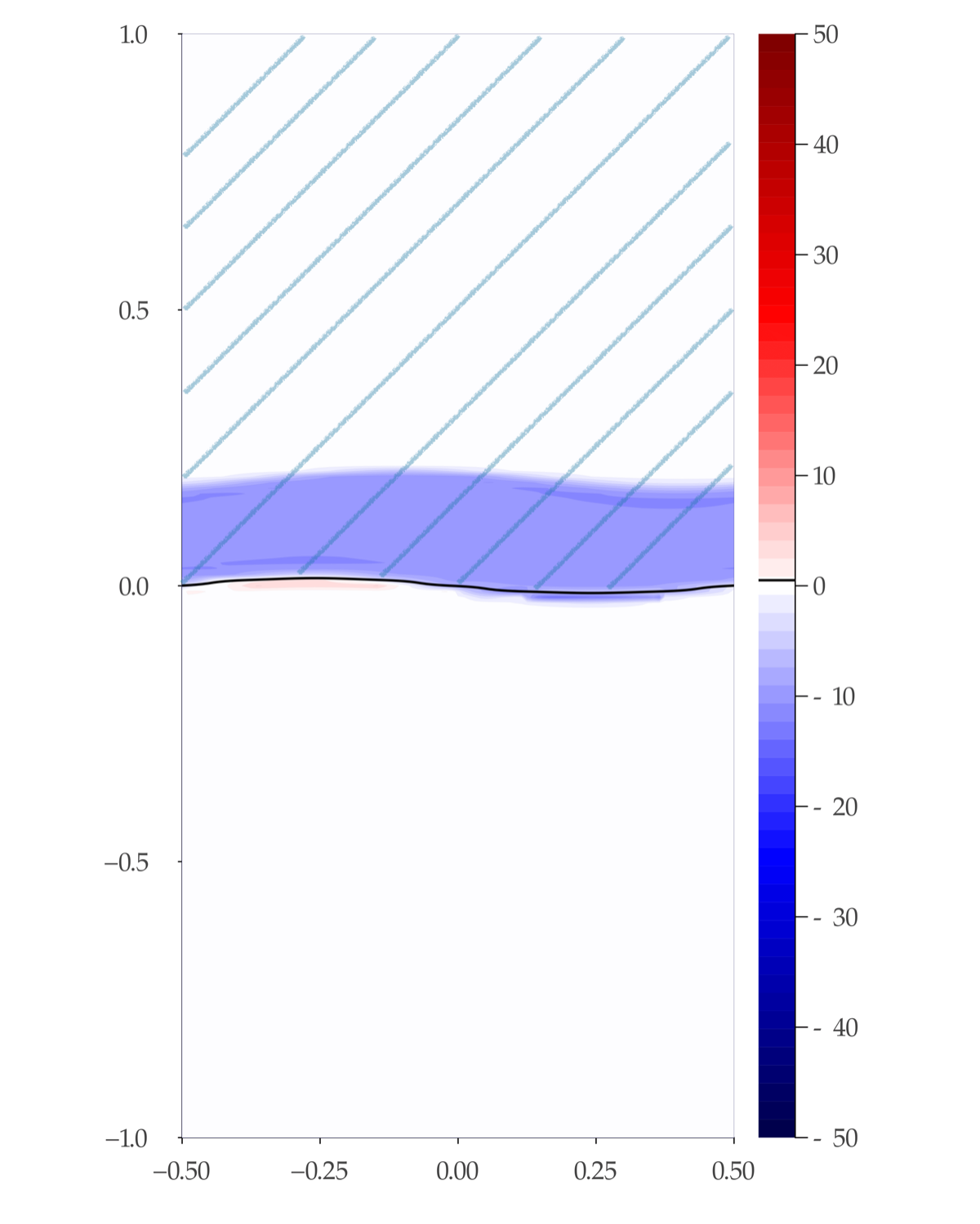}
        \caption{$tU/L=2.0$.}
    \end{subfigure}  
    \hfill
    \begin{subfigure}[b]{0.18\textwidth}
        \centering
        \adjincludegraphics[trim={{.18375\width} {.055\height} {.24\width} {.015\height}}, 
                 clip, height=0.28\textheight]{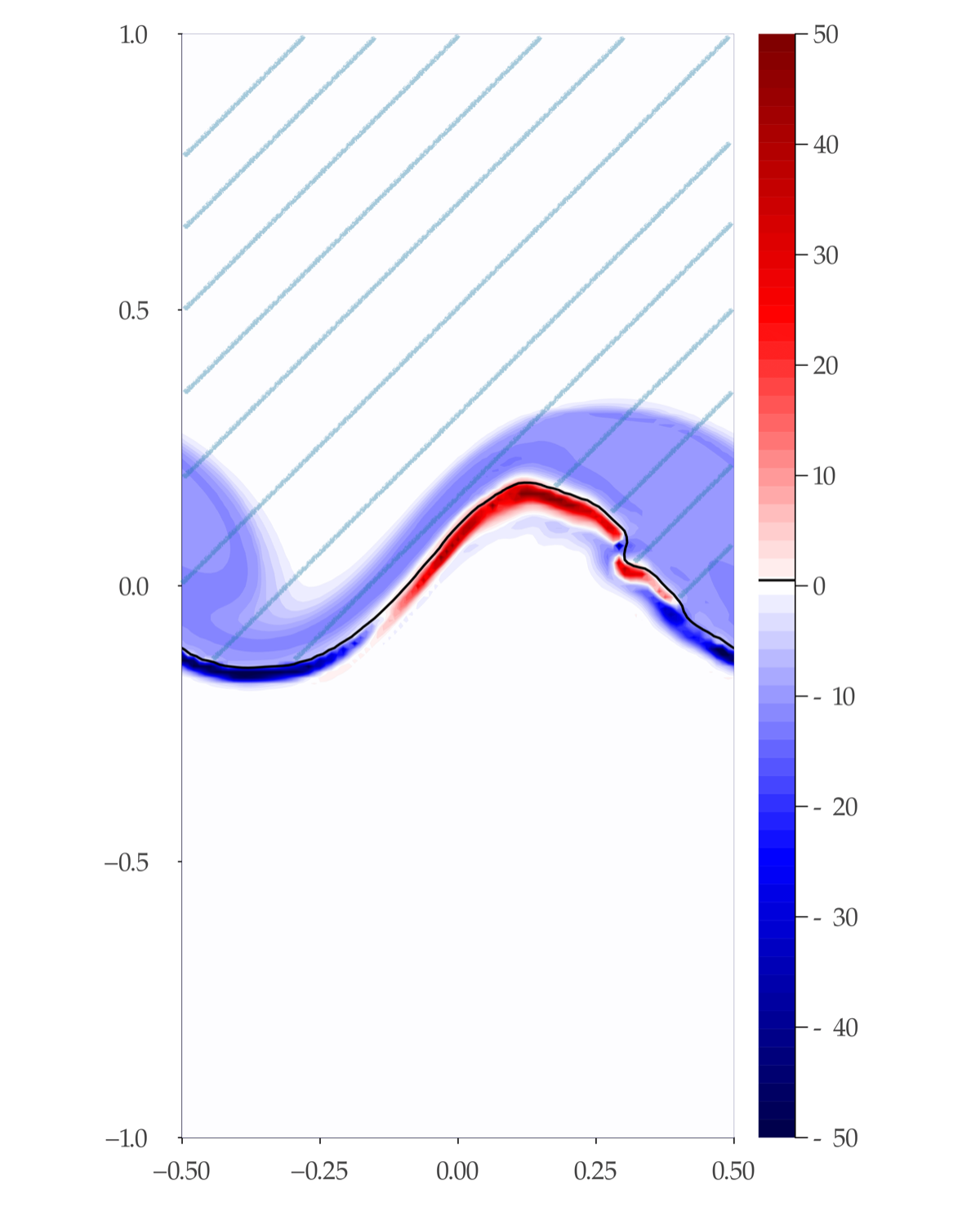}
        \caption{$tU/L=3.0$.}
    \end{subfigure}  
    \hfill
    \begin{subfigure}[b]{0.18\textwidth}
        \centering
        \adjincludegraphics[trim={{.18375\width} {.055\height} {.24\width} {.015\height}}, 
                 clip, height=0.28\textheight]{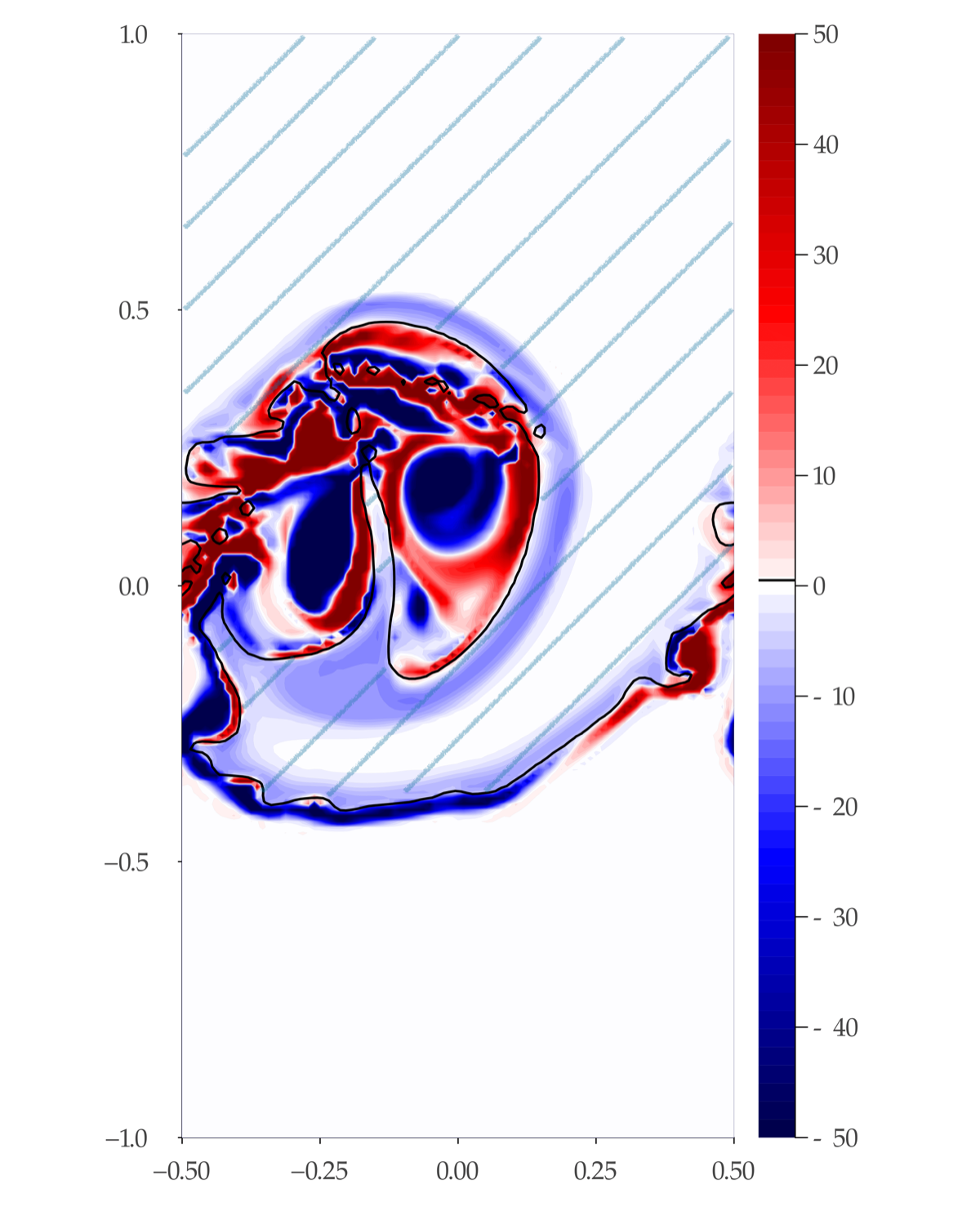}
        \caption{$tU/L=3.8$.}
    \end{subfigure}
    \hfill
    \begin{subfigure}[b]{0.26\textwidth}
        \centering
        \adjincludegraphics[trim={{.18375\width} {.055\height} {0} {.015\height}}, 
                clip, height=0.28\textheight]{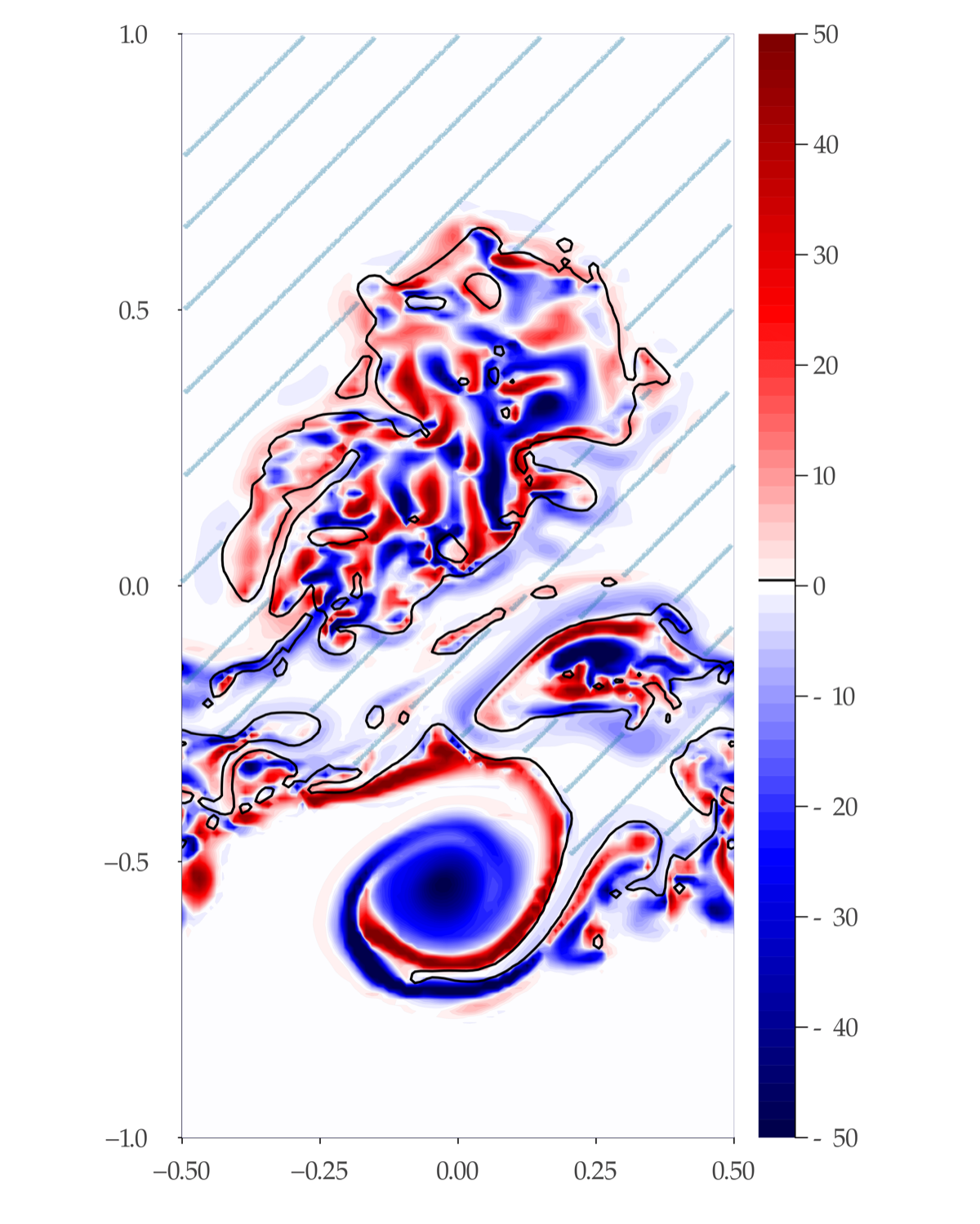}
        \caption{$tU/L=5.0$.}
    \end{subfigure}
    
    \caption{Evolution of thick shear layer Kelvin-Helmholtz instability on the interface, contoured by non-dimensional vorticity. The black line is the interface. Teal blue hatches signify the water parts. Green arrows show the initial horizontal velocity profile. $(N_x,N_y) = (96,192)$. $tU/L=3.8$ in panel (d) is approximately when the maximum kinetic energy drain happens.}
    \label{fig:KH}
\end{figure*}

\begin{figure}
    \centering
    \includegraphics[width=\linewidth]{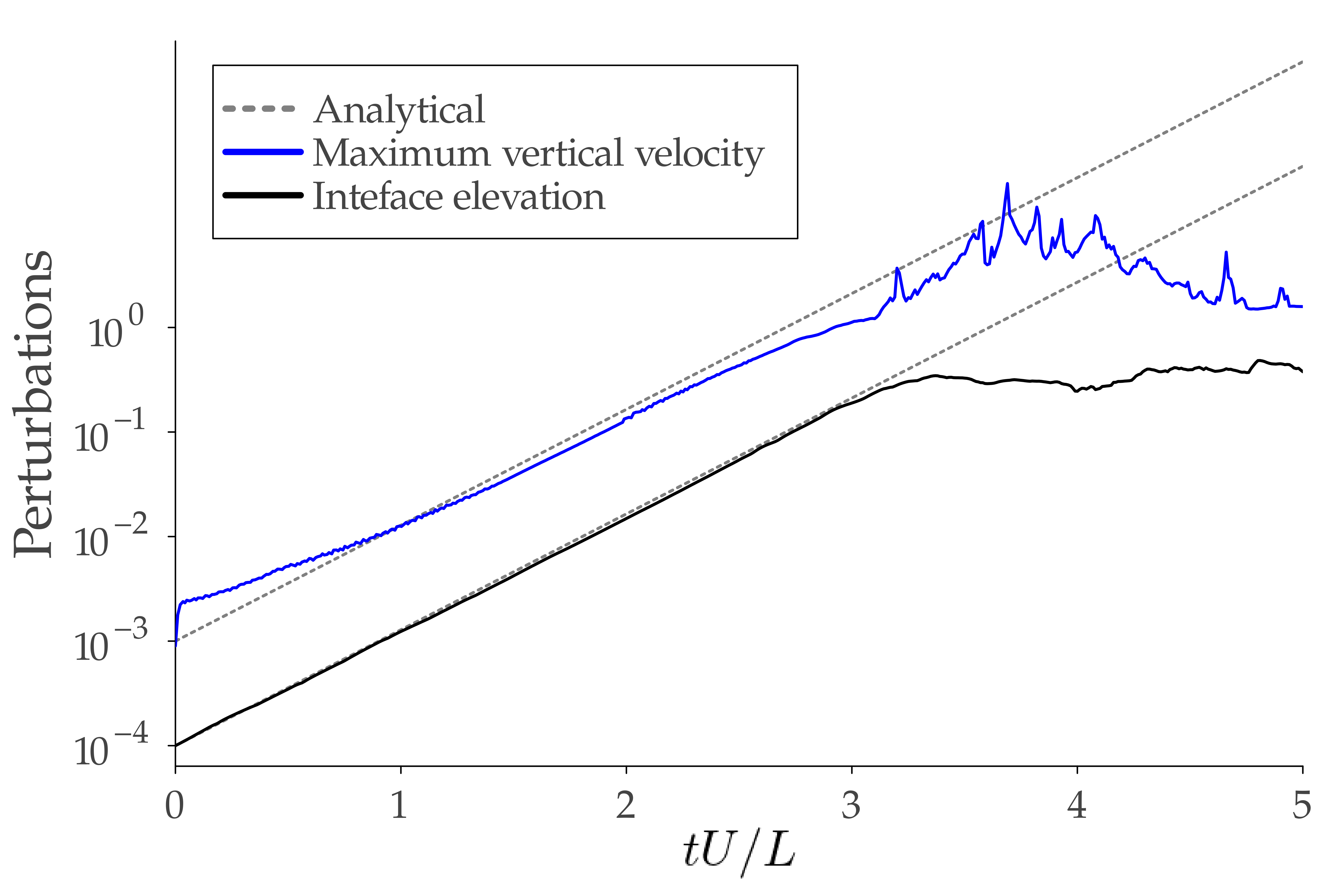}
    \caption{Temporal evolution of perturbations of interface amplitude and maximum vertical velocity in Kelvin-Helmholtz instability, where the dashed line denotes the theoretical prediction. The interface amplitude is filtered by Fourier transform such that it only contains the largest wavelength, \ie horizontal domain size.}
    \label{fig:KHvMax}
\end{figure}

\begin{figure}
    \centering
    \includegraphics[width=\linewidth]{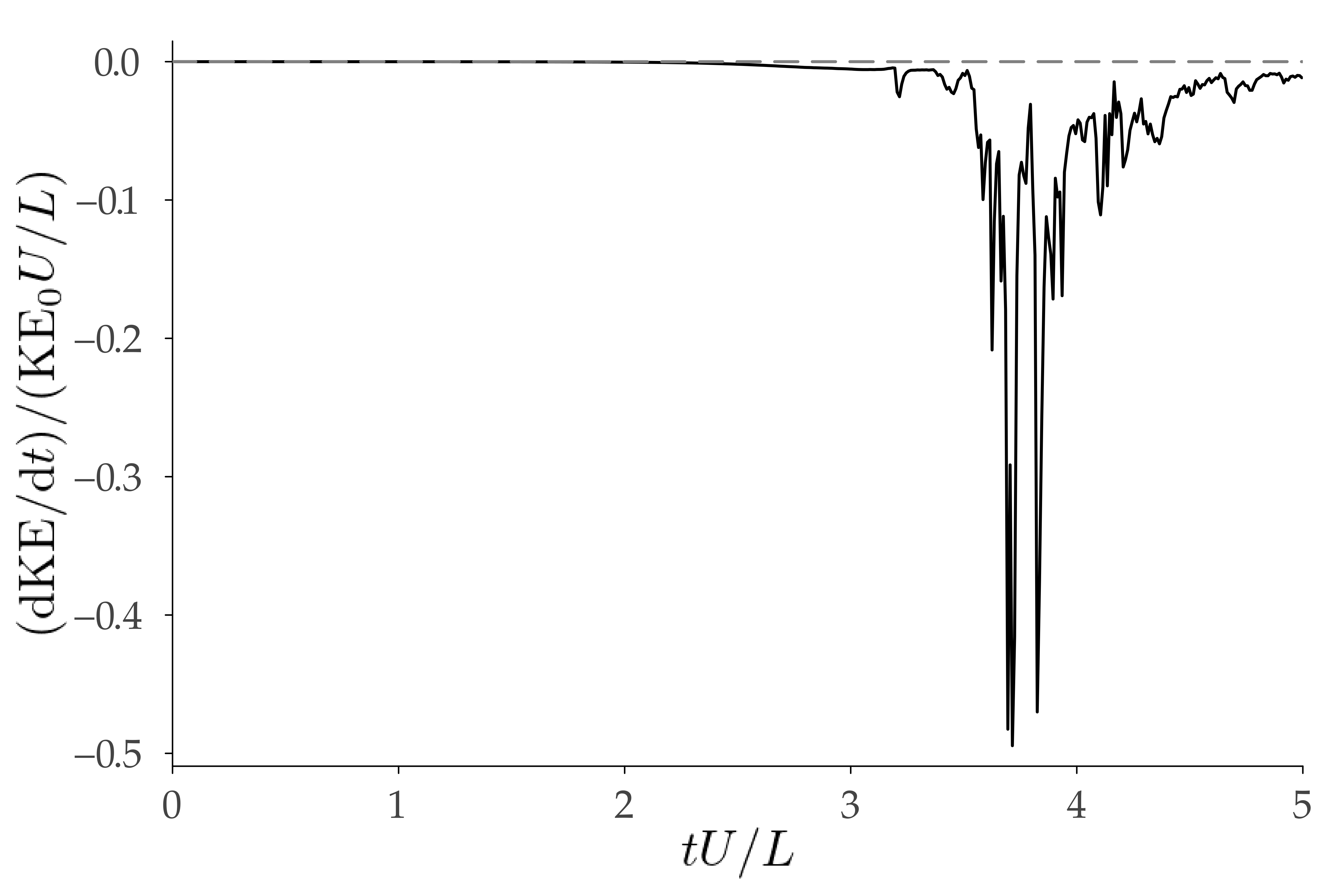}
    \caption{Time derivatives of total kinetic energy of Kelvin-Helmholtz test case with zero line indicated.}
    \label{fig:KHdKEdt}
\end{figure}

\begin{figure}
    \centering
    \includegraphics[width=\linewidth]{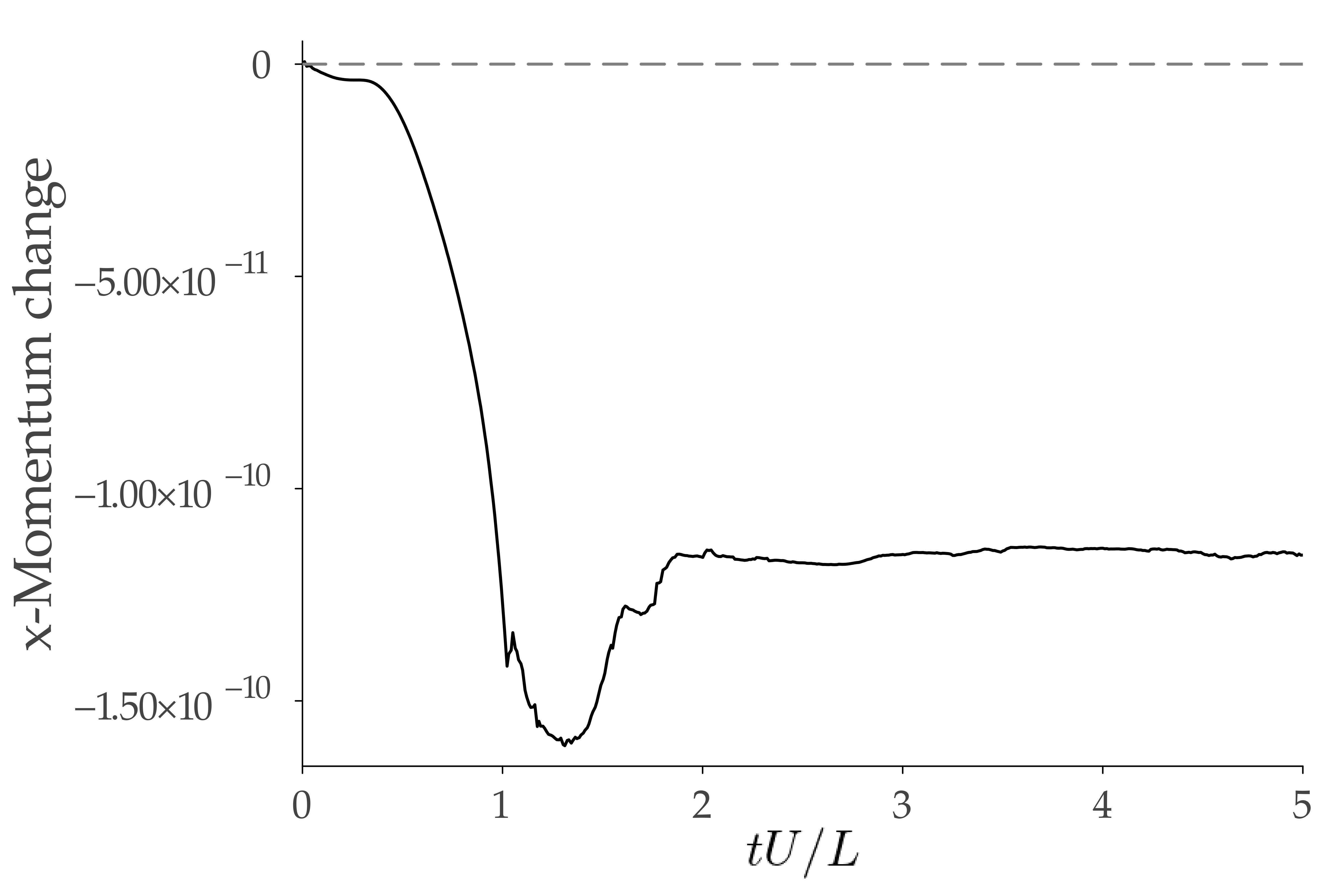}
    \caption{Change of the horizontal momentum of the Kelvin-Helmholtz case. The vertical momentum is not shown as the slip boundary condition is imposed on the top and bottom boundaries.}
    \label{fig:KHdqx}
\end{figure}

The Kelvin-Helmholtz instability provides a physically meaningful test for assessing whether the proposed numerical method can correctly capture the growth of physical perturbations without introducing artificial damping or numerical instability.
In particular, this case examines the ability of the scheme to preserve linear instability mechanisms while remaining robust during the subsequent nonlinear evolution.

The configuration follows that of \citet{2021Arrufat_MassMomentumConserv}.
A density ratio of $\lambda_\rho = 10^{-3}$ is selected to represent an air-water system.
The computational domain has dimensions $(L,2L)$, with the dense fluid initially occupying the upper half of the domain and the lighter fluid below, separated by a flat interface located at mid-height, as illustrated in \cref{fig:KH}a.
Gravity is omitted in order to isolate Kelvin-Helmholtz dynamics and exclude Rayleigh-Taylor effects.
Viscosity is also neglected to directly assess the robustness of the inviscid solver.

A finite-thickness shear layer is imposed in the dense phase.
The horizontal velocity transitions smoothly from $(1,0)U$ to $(-1,0)U$, as indicated by the green arrows in \cref{fig:KH}a, creating a velocity deflection near the interface that drives the instability.
Monochromatic perturbations with wavelength $L$ are applied simultaneously to the interface position and the vertical velocity field.
The vertical velocity perturbations decay exponentially away from the interface.
The grid resolution is $(N_x, N_y) = (96, 192)$, with periodic boundary conditions in the horizontal ($x$) direction and slip-wall boundary conditions in the vertical ($y$) direction.

During the early stages of the simulation, both the interface elevation and the maximum vertical velocity exhibit clear exponential growth in agreement with theoretical predictions for Kelvin-Helmholtz instability (\cref{fig:KHvMax}).
The smooth and monotonic evolution of the interface amplitude confirms the correct excitation and preservation of the prescribed monochromatic mode, which is also visually evident in \cref{fig:KH}a-c.
Minor oscillations observed in the early-time evolution of the vertical velocity are attributed to transient adjustment of the flow from the prescribed initial condition and do not affect the overall growth rate.

At later times, the flow enters a strongly nonlinear regime characterized by energetic interface roll-ups, vortex pairing, and eventual collapse of multiphase structures (\cref{fig:KH}d,e).
This stage used to pose a significant challenge to numerical stability due to the presence of intense localized velocities and sharp interfacial deformation.
Despite these demanding conditions, the proposed method successfully maintains bounded total kinetic energy throughout the simulation.
Energy dissipation occurs only after the flow becomes irregular and multiscale, consistent with the iLES philosophy (\cref{fig:KHdKEdt}).
Moreover, even in the presence of these fierce nonlinear phenomena, the method effectively confines the high-velocity regions to the light-fluid (air) phase, without spurious penetration of extreme velocities into the dense phase.
Throughout the simulation, horizontal momentum is conserved to solver tolerance, as shown in \cref{fig:KHdqx}, further confirming the consistency and robustness of the proposed formulation.

\section{Special treatment of the viscous term}

\subsection{Discretization scheme}

\begin{figure}
    \centering
    \includegraphics[width=\linewidth]{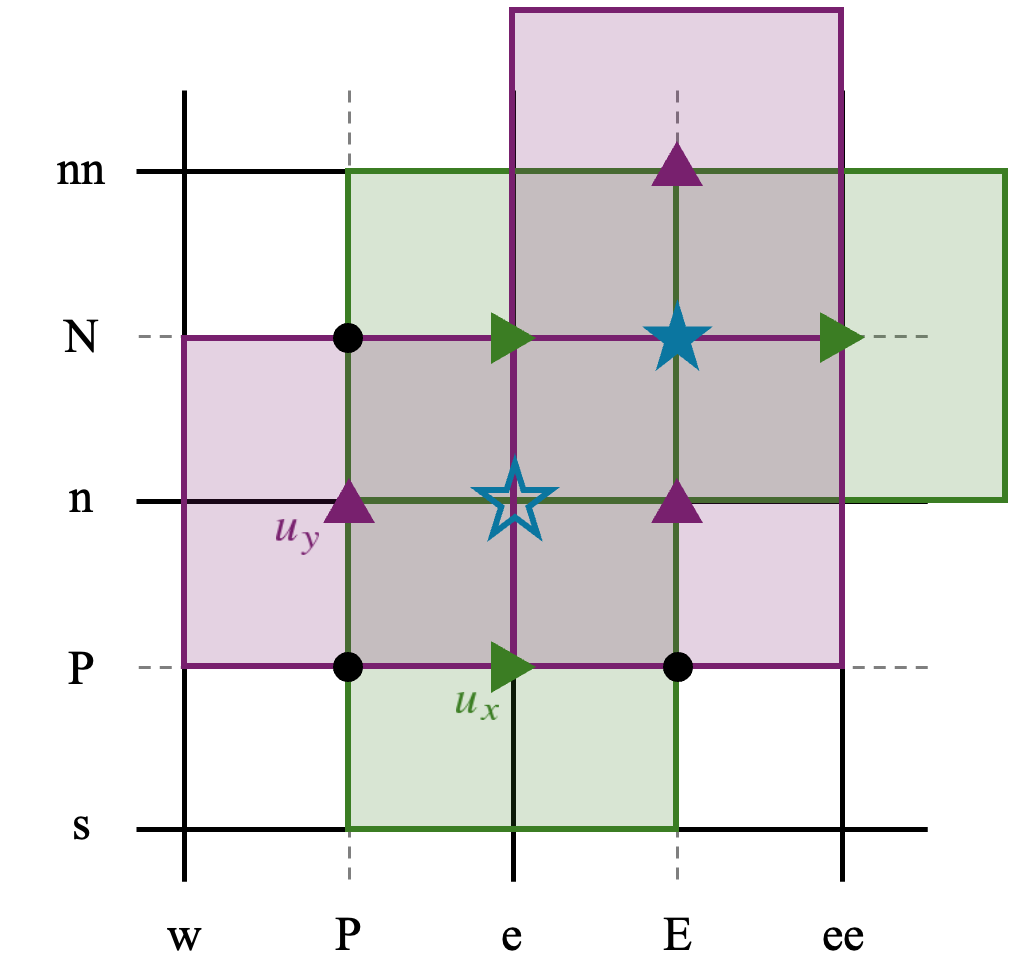}
    \caption{Locations at which viscosity is evaluated in a staggered-grid configuration. The hollow star denotes the collocated-cell vertex (edge in 3D), used for cross-difference terms, while the solid star denotes the collocated-cell center, used for inline-difference terms.}
    \label{fig:StaggeredGridViscous}
\end{figure}

The viscous term is discretized following the standard bi-rotational formulation
\citep{2015Campbell_Dissertation,2011Tryggvason_DNS}.
Shear-stress momentum fluxes are evaluated at the faces of staggered momentum cells
(\cref{fig:StaggeredGridViscous}).
The viscous contribution is naturally decomposed into cross-difference
($\partial_j u_i,\, i\neq j$) and inline-difference ($\partial_j u_i,\, i=j$) components.
Accordingly, dynamic viscosity must be evaluated at two distinct locations:
collocated-cell vertices (hollow stars) for cross-difference terms and
collocated-cell centers (solid stars) for inline-difference terms.

The viscosity should be interpolated from the surrounding staggered values.
Common choices include arithmetic and harmonic averages; for example, the cross-difference term at vertex $\mathrm{ne}$,
\begin{align*}
    \mu_{\mathrm{ne;arithmetic}} &= \frac{\mu_\mathrm{n} + \mu_\mathrm{nE} + \mu_\mathrm{e} + \mu_\mathrm{Ne}}{4}, \\
    \mu_{\mathrm{ne;harmonic}} &= \frac{4}{\mu_\mathrm{n}^{-1}+\mu_\mathrm{nE}^{-1}+\mu_\mathrm{e}^{-1}+\mu_\mathrm{Ne}^{-1}} .
\end{align*}
To update velocity, the viscous flux, thus $\mu$, is divided by the density at the staggered-cell center.
Near material interfaces, however, this division can become unbounded due to mismatched density and viscosity interpolations.
This issue is intrinsic to both velocity and momentum formulations.

\subsection{Failure of pure interpolation schemes}

\begin{figure*}
    \centering
    \begin{subfigure}[t]{0.49\linewidth}
        \centering
        \includegraphics[width=0.7\textwidth]{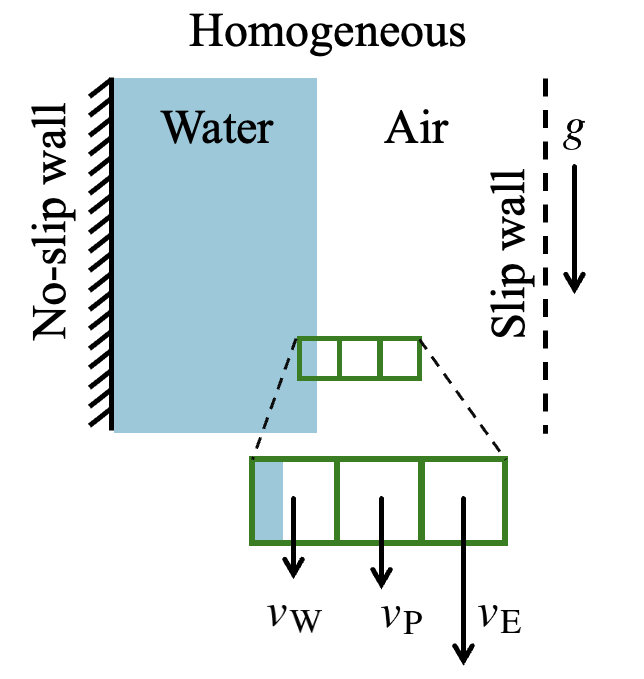}
        \caption{Stratified Poiseuille flow driven by a horizontal body force.}
        \label{fig:Poiseuille}
    \end{subfigure}
    \hfill
    \begin{subfigure}[t]{0.49\linewidth}
        \centering
        \includegraphics[width=0.9\textwidth]{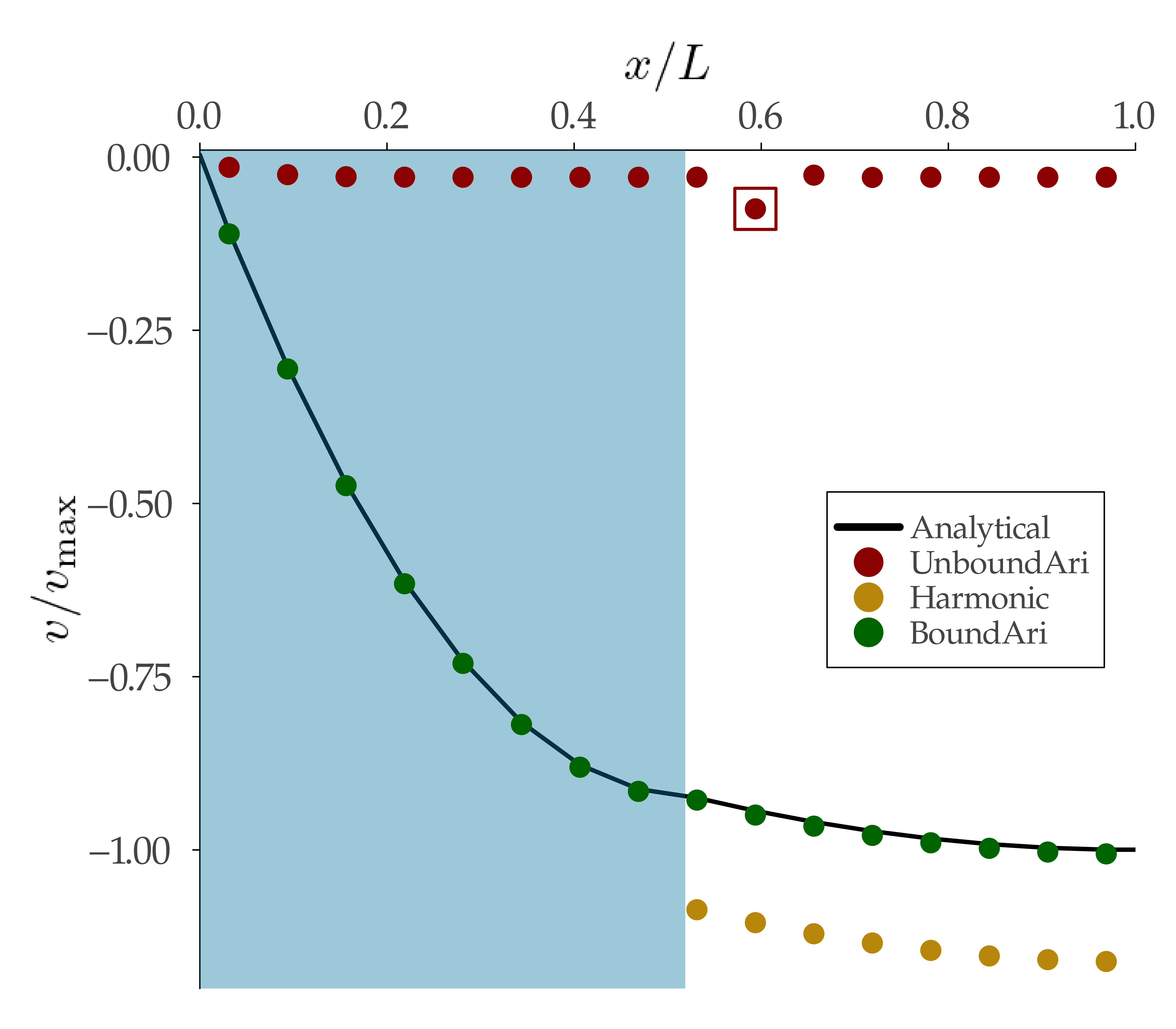}
        \caption{Velocity profiles of stratified Poiseuille flow.}
        \label{fig:AriHar16_FinalU}
    \end{subfigure}
    \caption{Gravity-driven stratified Poiseuille flow--case setup and simulation results. \textit{UnboundAri}: arithmetic averaging without bounded treatment; \textit{Harmonic}: harmonic mean following \citet{2011Tryggvason_DNS}; \textit{BoundAri}: arithmetic averaging with bounded treatment. The horizontal line marks the interface. The dark red box indicates the onset of divergence for \textit{UnboundAri}. $(N_x,N_y)=(16,1)$.}
    \label{fig:Poiseuilleall}
\end{figure*}

To illustrate the limitation of standard interpolation schemes, we consider a quasi-one-dimensional stratified Poiseuille flow driven downward by gravity, as shown in \cref{fig:Poiseuille}.
The water occupies the region near the no-slip wall, while the air contacts a slip-wall boundary.
The vertical direction is considered homogeneous as a quasi-one-dimensional case.
The material properties are $\rho_\mathrm{air}/\rho_\mathrm{water}=10^{-3}$, $\mu_\mathrm{air}/\mu_\mathrm{water}=10^{-2}$, and thus $\bm{\nu}_\mathrm{air}/\bm{\nu}_\mathrm{water}=10$.\footnote{The kinematic viscosity $\bm{\nu}$ is typeset in boldface to distinguish it from the vertical velocity $v$; this notation does not imply a vector quantity.}

The resulting velocity profiles are shown in \cref{fig:AriHar16_FinalU}.
The arithmetic average leads to numerical divergence within a few time steps.
Although the harmonic mean avoids divergence, it significantly over-predicts the velocity gradient, reflecting its excessive bias toward the low-viscosity phase and its neglect of inertial effects.

The root cause of both issue lies in the lack of boundedness of the \textit{kinematic} viscosity, which directly controls viscous CFL stability \citep{2015Campbell_Dissertation}.
Consider the control volume $\mathrm{P}$ immediately next to the interface in \cref{fig:Poiseuille}.
The semi-discrete viscous contribution reads
\begin{align*}
    \partial_t v_\mathrm{P} =
    \frac{
        \mu_\mathrm{e}(v_\mathrm{E}-v_\mathrm{P})
        - \mu_\mathrm{w}(v_\mathrm{P}-v_\mathrm{W})
    }{\rho_\mathrm{P}} - g .
\end{align*}
Two effective kinematic viscosities appear: $\mu_\mathrm{e}/\rho_\mathrm{P}$ and $\mu_\mathrm{w}/\rho_\mathrm{P}$.
While the eastern contribution remains bounded, the western contribution involves dividing a partially water-weighted dynamic viscosity by the air density, which can be excessively large.
For instance, with a $30\%$ water volume fraction in the cell W,
\begin{align*}
    \text{Arithmetic: }\quad
    \frac{\mu_\mathrm{w}}{\rho_\mathrm{P}} &=
    \frac{0.15\mu_\mathrm{water}+0.85\mu_\mathrm{air}}{\rho_\mathrm{air}}
    = 15.85\,\bm{\nu}_\mathrm{air}, \\
    \text{Harmonic: }\quad
    \frac{\mu_\mathrm{w}}{\rho_\mathrm{P}} &=
    \frac{\frac{2}{\frac{1}{0.3\mu_\mathrm{water}+0.7\mu_\mathrm{air}}+\frac{1}{\mu_\mathrm{air}}}}{\rho_\mathrm{air}}
    = 1.9\,\bm{\nu}_\mathrm{air}.
\end{align*}
Both exceed the largest physical kinematic viscosity, $\bm{\nu}_\mathrm{air}$ here, violating the viscous CFL constraint and leading to instability or loss of accuracy.

\subsection{Bounded kinematic viscosity}

To guarantee numerical stability, the effective kinematic viscosity must remain bounded.
We enforce this requirement by limiting the dynamic viscosity.
At vertex $\mathrm{ne}$ in \cref{fig:StaggeredGridViscous}, which is shared by four staggered momentum cells, we define
\begin{align}
    \mu_\mathrm{ne}
    = \min\!\left(\mu_{\mathrm{ne;orig}},\,
    \rho_{\mathrm{ne;min}}\,\bm{\nu}_{\mathrm{ne}}\right),
\end{align}
where $\mu_{\mathrm{ne;orig}}$ is the original interpolated viscosity (arithmetic, harmonic, etc.), and $\rho_{\mathrm{ne;min}}$ is the minimum density among the four associated momentum cells,
\begin{align*}
    \rho_{\mathrm{ne;min}}
    = \min_{\mathcal{C}\in\{\mathrm{n,nE,e,Ne}\}} \rho_{u;\mathcal{C}} .
\end{align*}
The upper bound $\bm{\nu}_{\mathrm{ne}}$ is chosen based on phase occupancy,
\begin{align*}
    \bm{\nu}_{\mathrm{ne}} =
    \begin{dcases}
        \bm{\nu}_\mathrm{water}, & \text{if } \sum_{\mathcal{C}\in\{\mathrm{n,nE,e,Ne}\}} f_\mathcal{C} \ge 2, \\
        \bm{\nu}_\mathrm{air}, & \text{otherwise}.
    \end{dcases}
\end{align*}
The inline contribution is also estimated with similar fashion but with only two staggered cell.

As shown in \cref{fig:AriHar16_FinalU}, the proposed limiter yields both stable and physically accurate solutions.
In the quasi-one-dimensional Poiseuille configuration, the viscosity at the western face naturally degenerates to the light-phase value, consistent with the physical configuration.
This bounded treatment therefore restores robustness without sacrificing accuracy.

\section{3D wave breakers}

\begin{figure*}[ht!]
    \centering
    \hfill
    \begin{subfigure}[t]{0.49\textwidth}
        \centering
        \includegraphics[width=\textwidth]{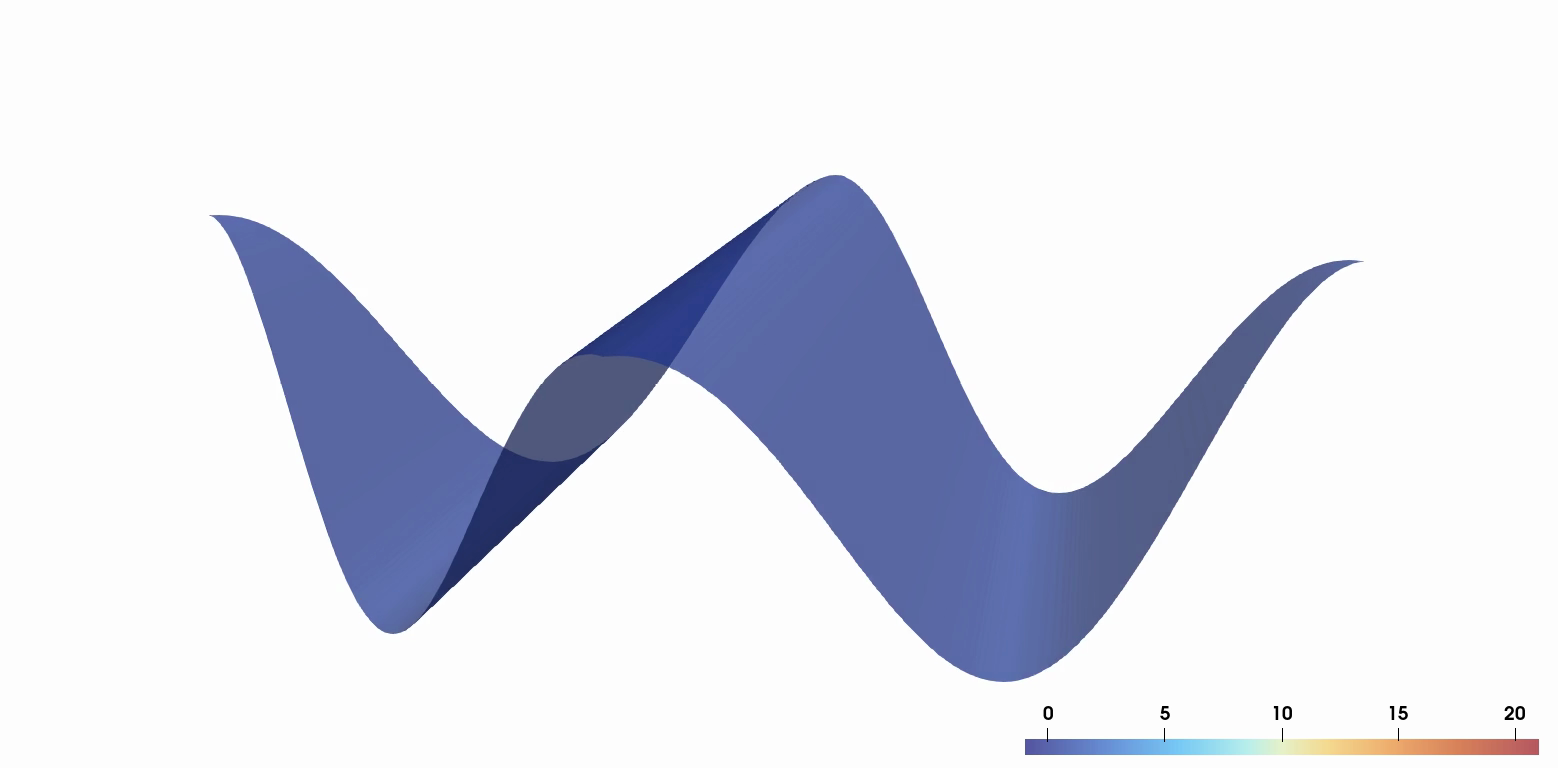}
        \caption{Initial condition, $tU/L=0.0$.}
    \end{subfigure}
    \hfill
    \begin{subfigure}[t]{0.49\textwidth}
        \centering
        \includegraphics[width=\textwidth]{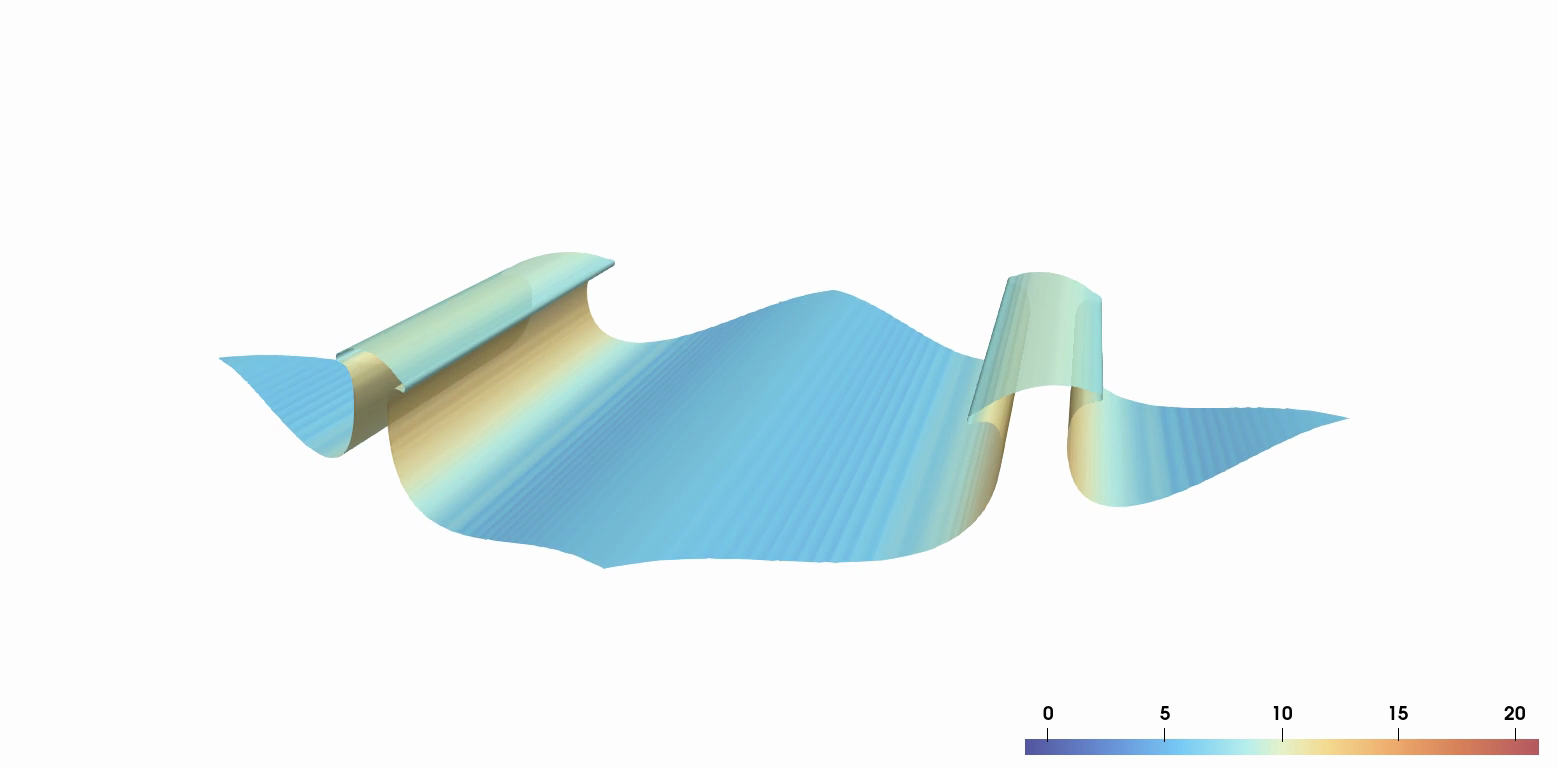}
        \caption{First kinetic-energy peak, $tU/L=0.22$.}
    \end{subfigure}
    \begin{subfigure}[t]{0.49\textwidth}
        \centering
        \includegraphics[width=\textwidth]{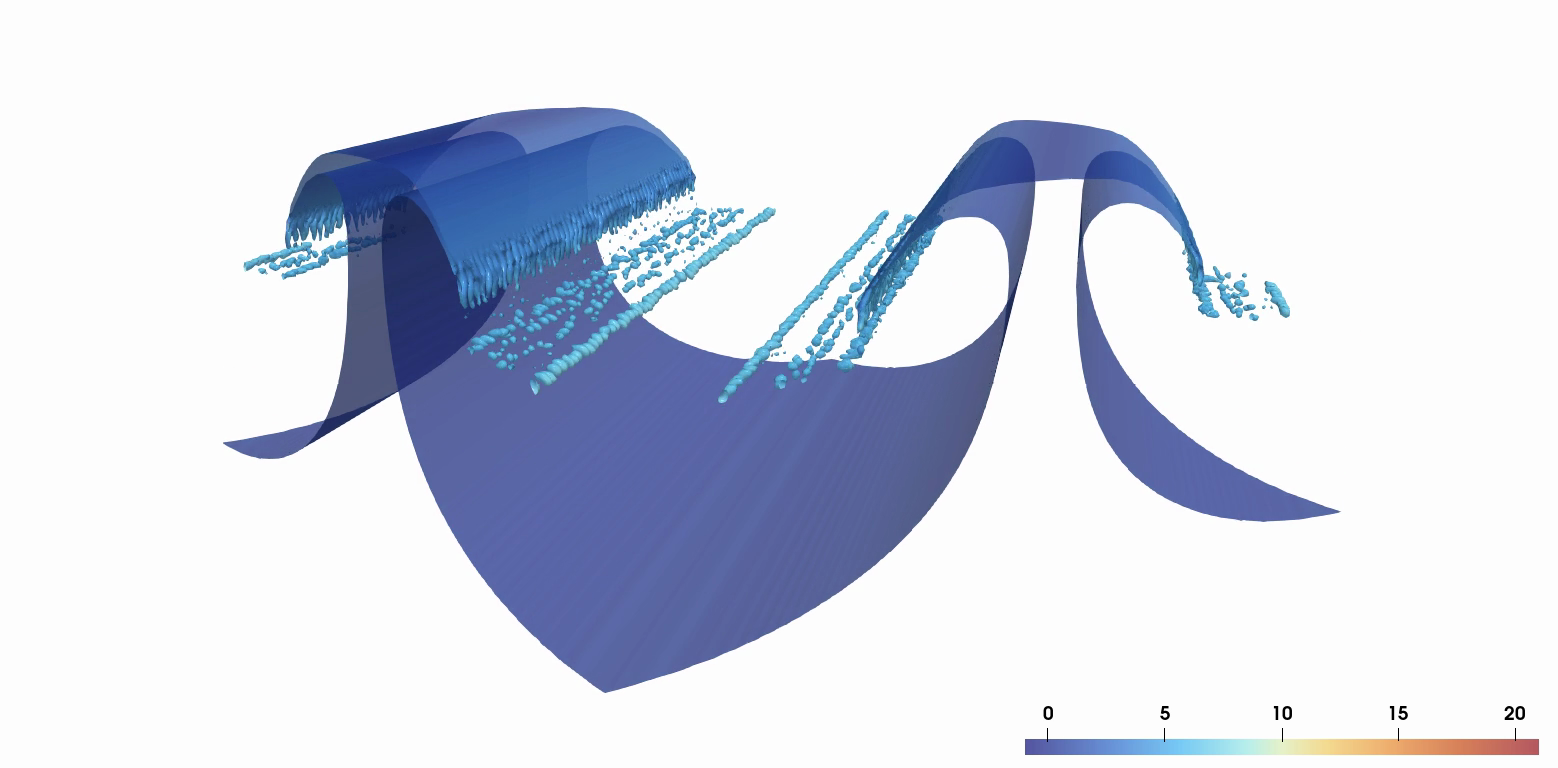}
        \caption{Second potential-energy peak, $tU/L=0.46$.}
    \end{subfigure}
    \hfill
    \begin{subfigure}[t]{0.49\textwidth}
        \centering
        \includegraphics[width=\textwidth]{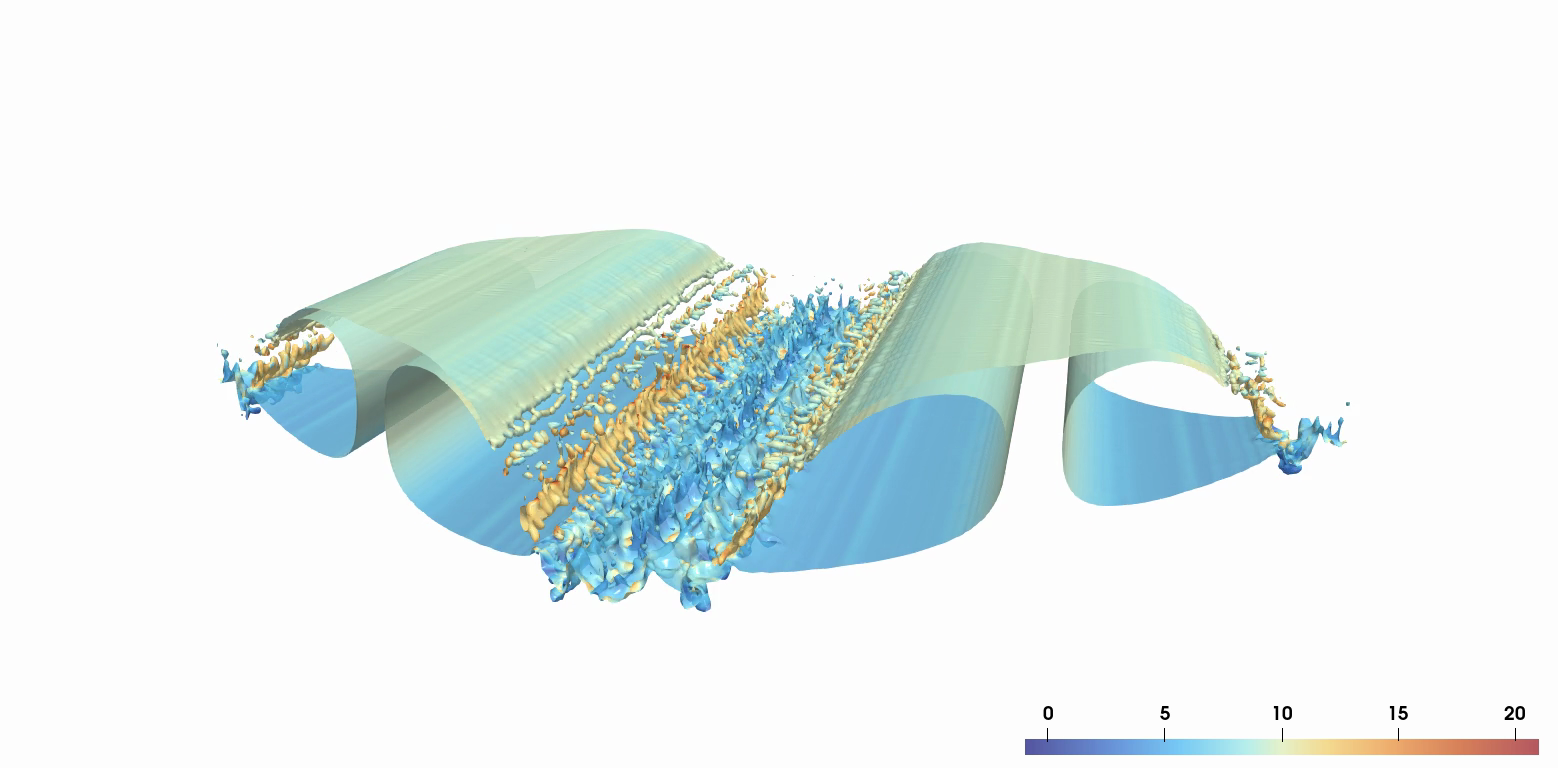}
        \caption{Second kinetic-energy peak and maximum energy dissipation, $tU/L=0.66$.}
    \end{subfigure}
    \begin{subfigure}[t]{0.49\textwidth}
        \centering
        \includegraphics[width=\textwidth]{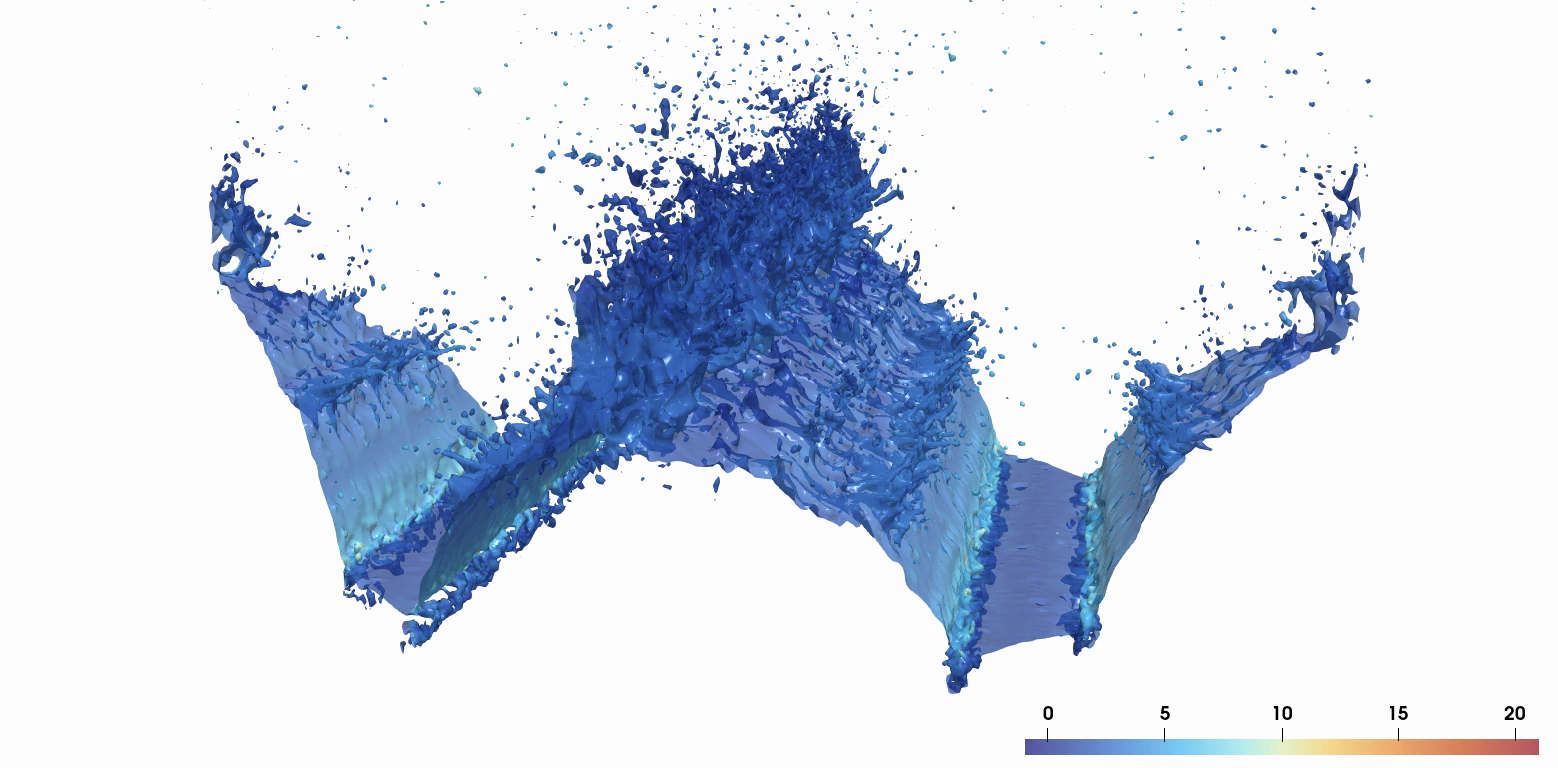}
        \caption{Third potential-energy peak, $tU/L=0.95$.}
    \end{subfigure}
    \hfill
    \begin{subfigure}[t]{0.49\textwidth}
        \centering
        \includegraphics[width=\textwidth]{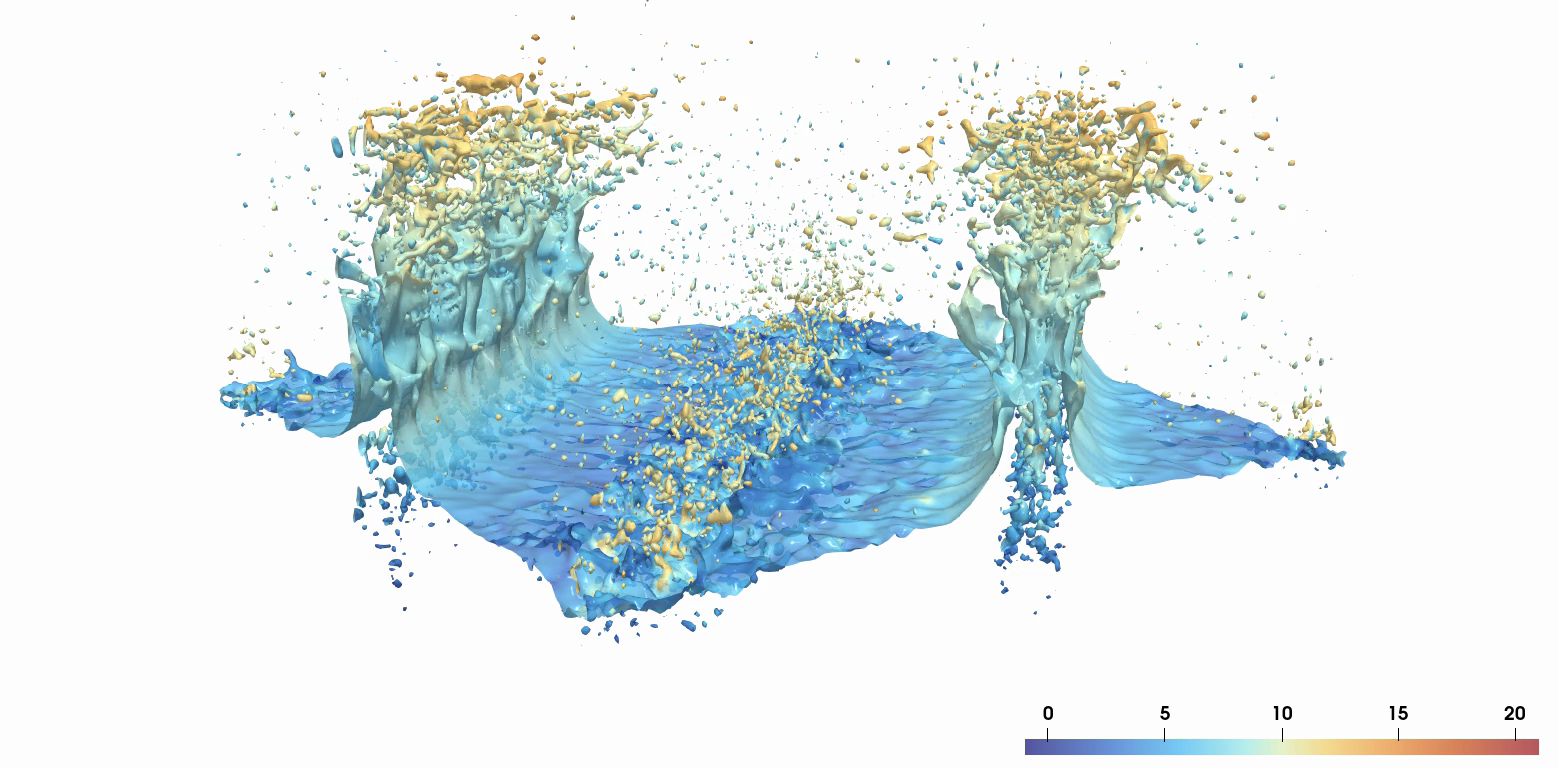}
        \caption{Third kinetic-energy peak, $tU/L=1.17$.}
    \end{subfigure}
    \caption{Evolution of a diagonal sine standing wave with inviscid condition. The free surface is shown and colored by velocity magnitude. Instants corresponding to peaks or troughs of kinetic and potential energy are presented. $N_x = N_y = N_z = 256$.}
    \label{fig:DiagSine256}
\end{figure*}

\begin{figure*}[ht!]
    \centering
    \hfill
    \begin{subfigure}[b]{0.49\textwidth}
        \centering
        \includegraphics[width=\textwidth]{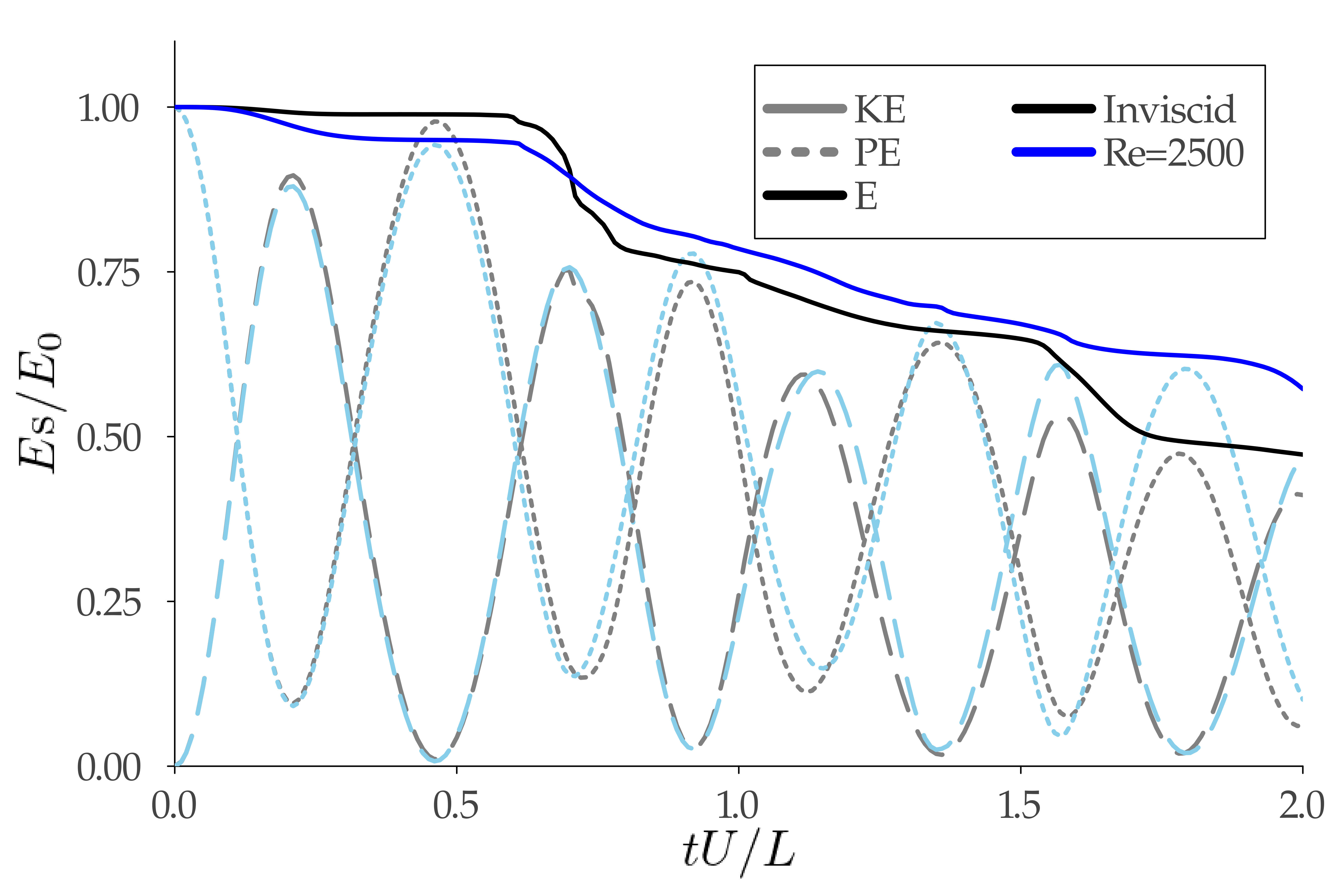}
        \caption{Total mechanical energy evolution.}
    \end{subfigure}
    \hfill
    \begin{subfigure}[b]{0.49\textwidth}
        \centering
        \includegraphics[width=\textwidth]{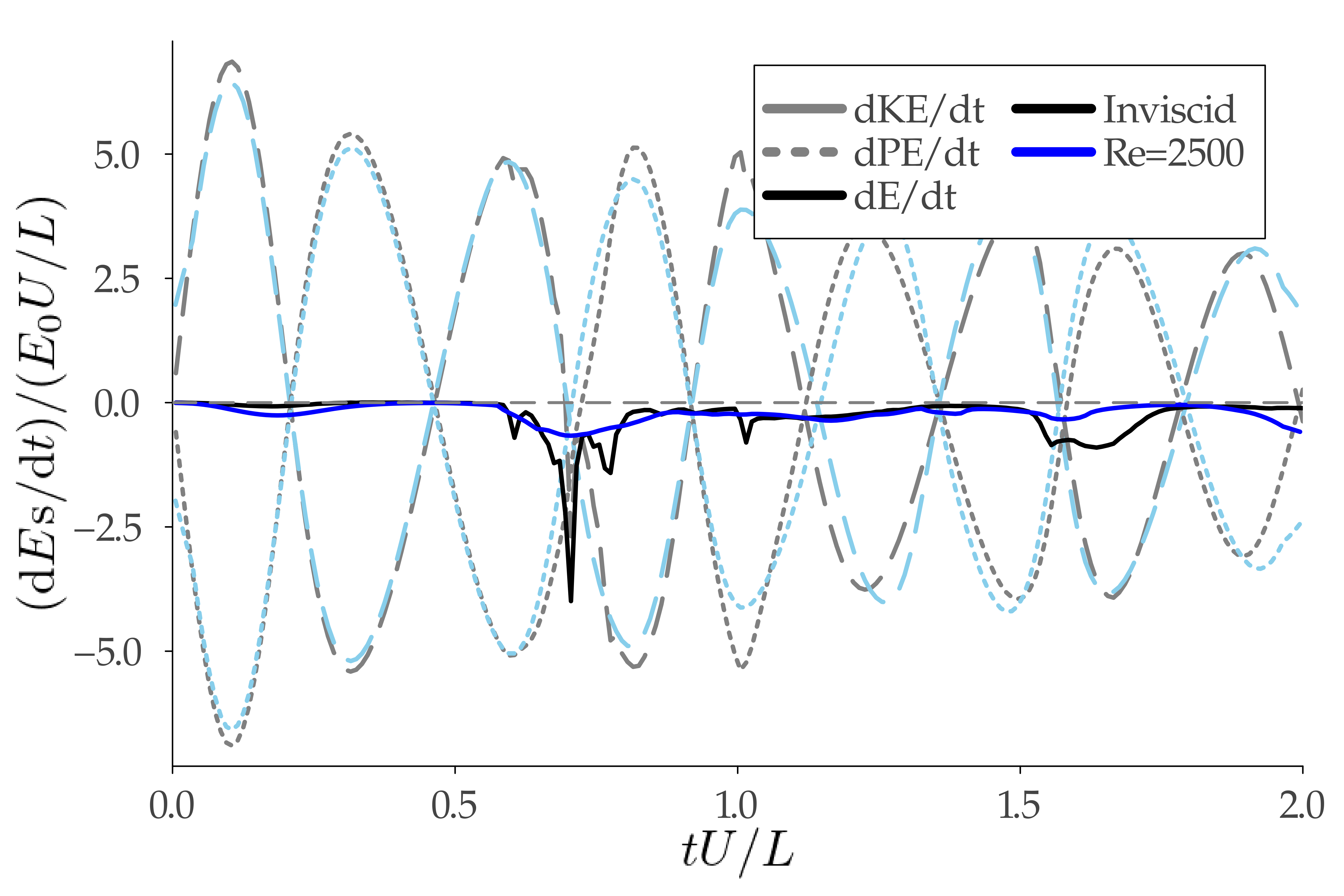}
        \caption{Temporal change of mechanical energy components.}
    \end{subfigure}
    \caption{Mechanical energy evolution and its temporal derivatives. The zero level of the energy derivative is indicated by a gray dashed line. The color family of black (resp.\ blue) represents inviscid (resp.\ viscous) solution.}
    \label{fig:sinewave_256ECompare}
\end{figure*}

\begin{figure}
    \centering
    \includegraphics[width=\linewidth]{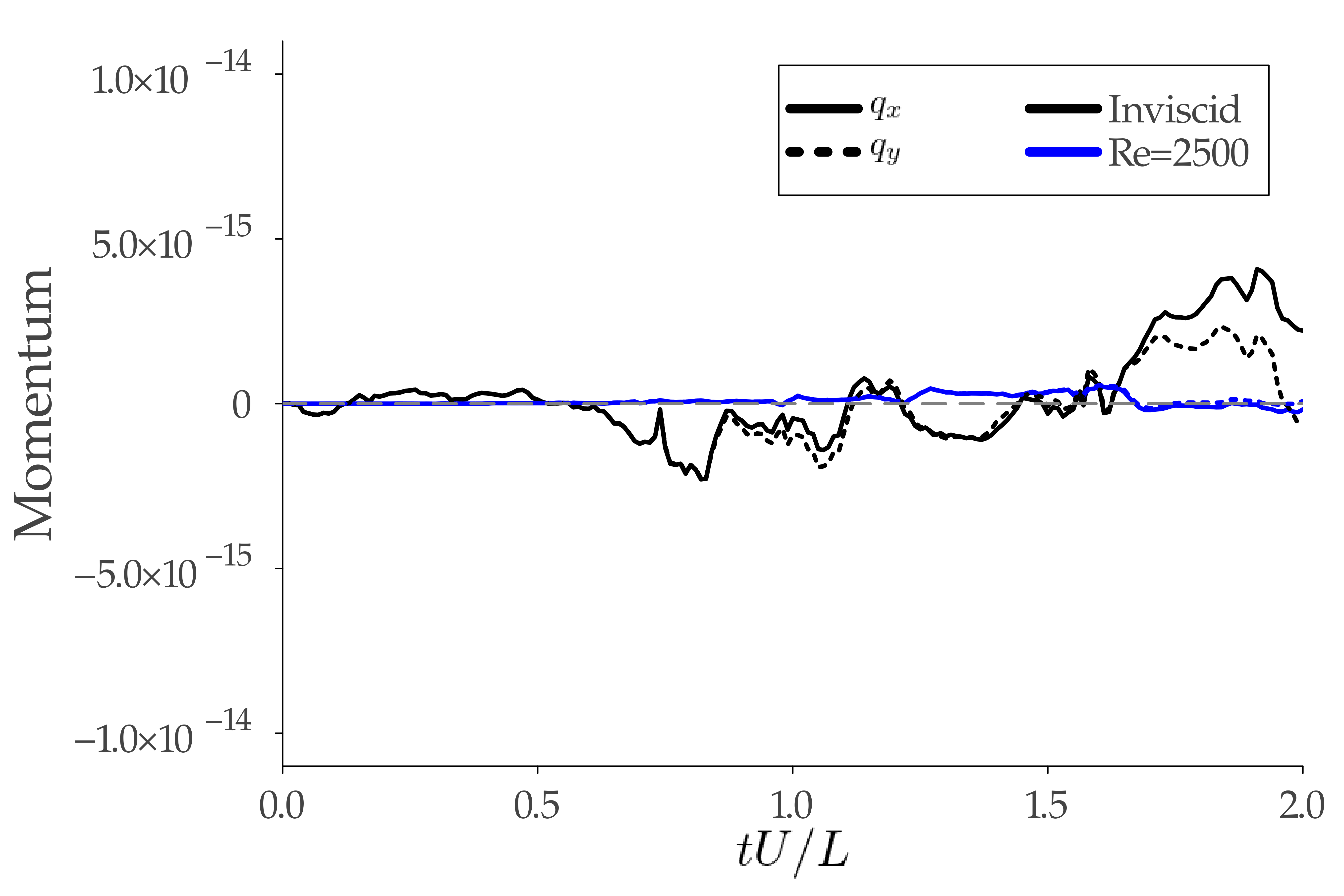}
    \caption{Relative change of horizontal momentum for the wave-breaker case. Vertical momentum is omitted due to slip boundary conditions at the top and bottom. The color family of black (resp.\ blue) represents inviscid (resp.\ viscous) solution.}
    \label{fig:sinewave_256Momentum}
\end{figure}

\begin{figure*}
    \centering
    \hfill
    \begin{subfigure}[b]{0.49\textwidth}
        \centering
        \includegraphics[width=\textwidth]{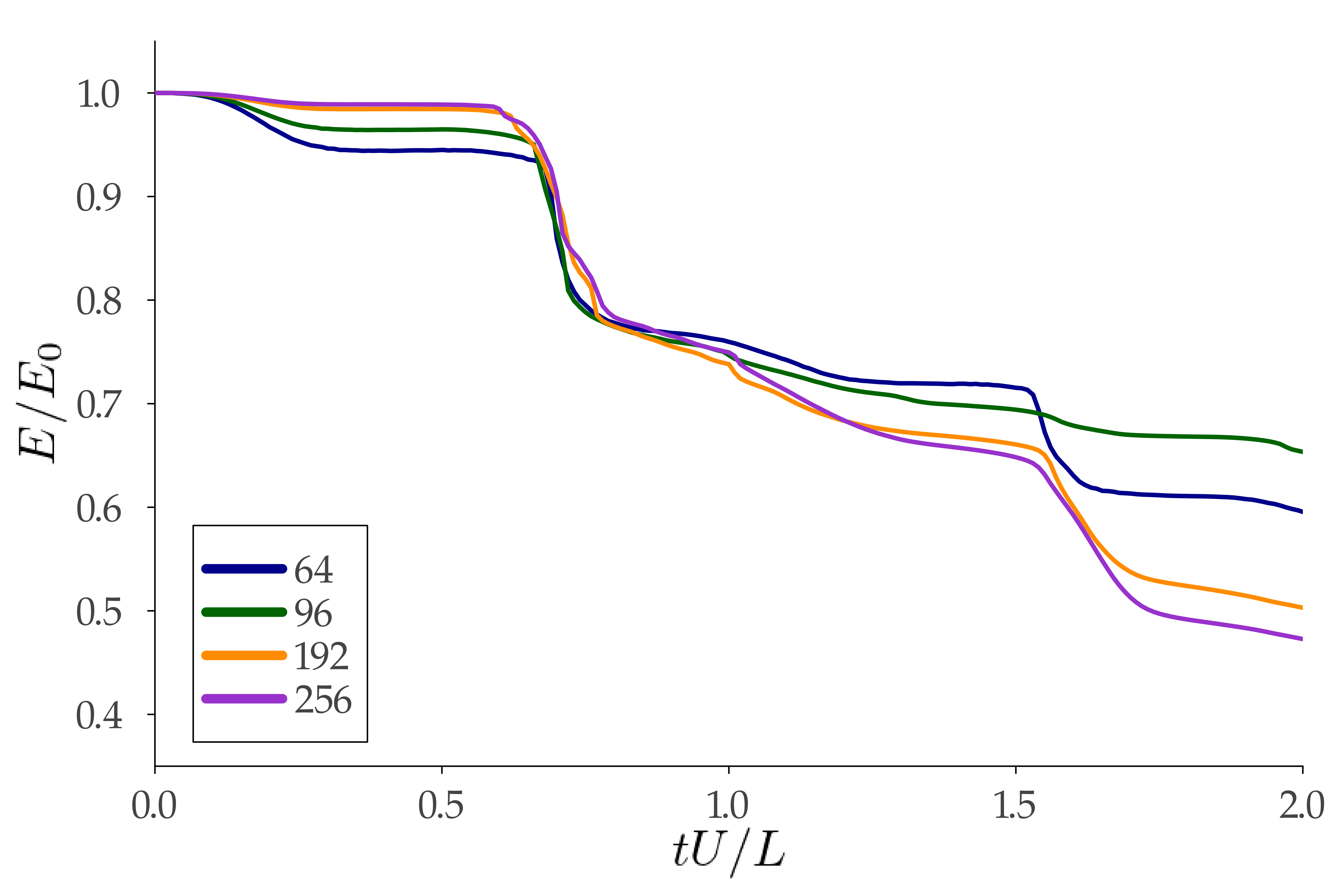}
        \caption{Total mechanical energy evolution.}
    \end{subfigure}
    \hfill
    \begin{subfigure}[b]{0.49\textwidth}
        \centering
        \includegraphics[width=\textwidth]{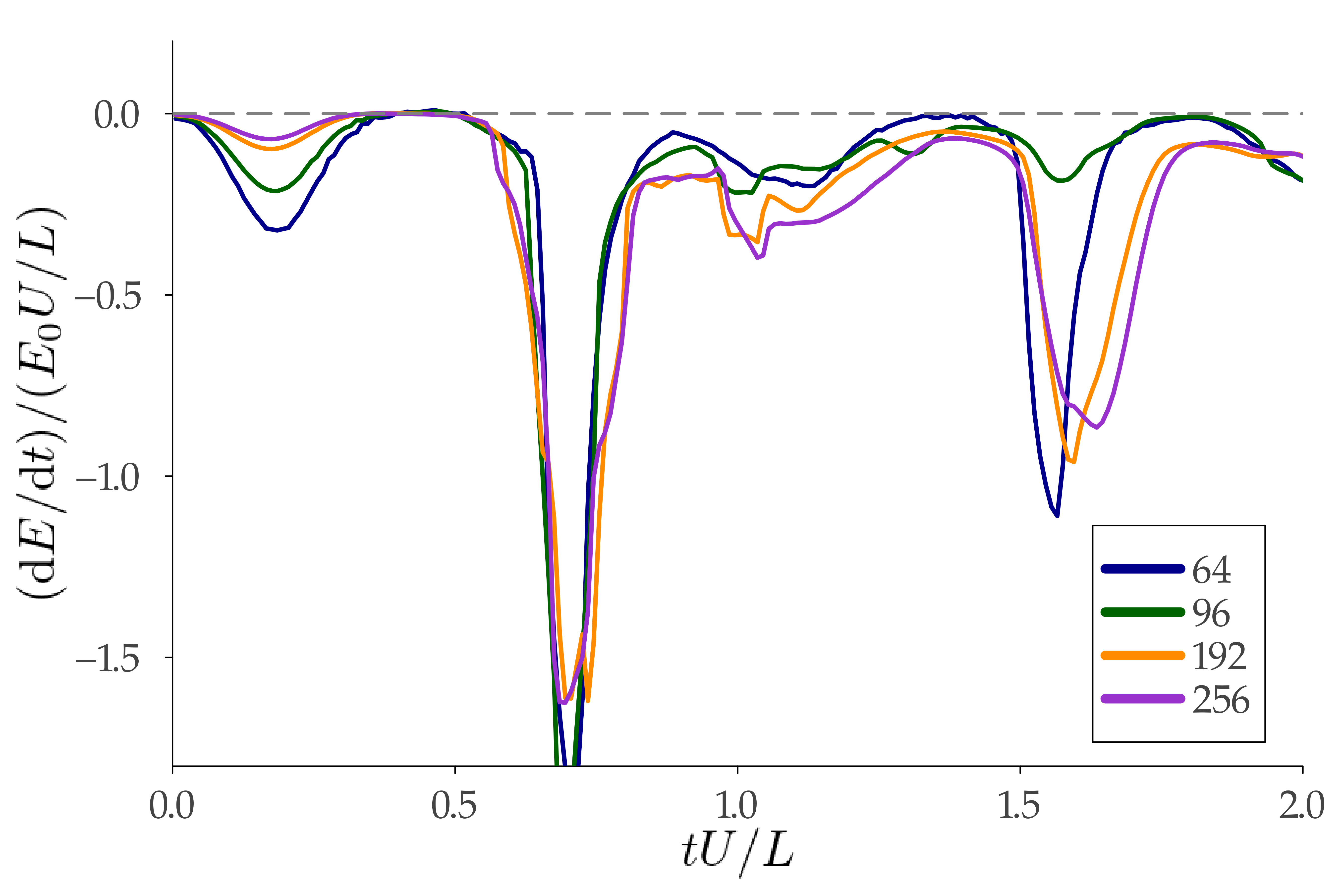}
        \caption{Temporal change of mechanical energy.}
    \end{subfigure}
    \caption{Grid-convergence of mechanical energy and its temporal derivative in inviscid simulation. The temporal derivative is window-averaged for clarity.}
    \label{fig:sinewave_ECompare}
\end{figure*}

The final test case considers a diagonal standing wave breaker, a canonical problem in naval hydrodynamics for which traditional velocity-based formulations are known to fail \citep{2010Fu_Favre}.
The imposed wave steepness is well beyond the linear regime, resulting in intense nonlinear interactions and strong concentration of kinetic energy near the air-water interface during wave overturning.

The initial free-surface elevation is prescribed as
\begin{align*}
z = 0.2L\sin\left( \frac{2\pi(x+y)}{L} \right),
\end{align*}
where $L$ denotes the domain length in all three spatial directions, and the mean water level is located at mid-height of the domain.
The diagonal orientation of the wave relative to the Cartesian grid is deliberately chosen to avoid grid-aligned bias and to challenge the numerical scheme under nontrivial interface advection.
Periodic boundary conditions are applied in the horizontal directions ($x$ and $y$), while slip-wall boundary conditions are imposed at the top and bottom boundaries.

The characteristic velocity scale is taken as the Airy phase velocity, $U=\sqrt{g\tanh(kh)/k}$, with the wavenumber defined as $k = 2\sqrt{2}\pi/L$.
Simulations are performed both with and without the viscous term.
For the viscous case, the Reynolds number based on the water kinematic viscosity is $Re = UL/\bm{\nu}_{\mathrm{water}} = 2500$, which is approximately the maximum Reynolds number that can be reliably resolved on the finest grid with $N_x = N_y = N_z = 256$.

The evolution of the free surface on the finest grid is demonstrated in \cref{fig:DiagSine256}, with the corresponding mechanical energy evolution presented in \cref{fig:sinewave_256ECompare}.
At early times, potential energy is rapidly converted into kinetic energy.
Due to the large wave steepness, the crests do not evolve smoothly but instead form vertically oriented jets (\cref{fig:DiagSine256}b).
These jets subsequently undergo bifurcation, transitioning first into plunging breakers and then into spilling breakers (\cref{fig:DiagSine256}c).
The spilling breakers interact strongly with the underlying free surface and are responsible for the most significant energy-drain events owing to their highly energetic impacts (\cref{fig:DiagSine256}d).
This stage is followed by intense air-water interaction, marked by vigorous entrainment and fragmentation (\cref{fig:DiagSine256}d-f).

Despite the extreme nonlinearity and energetic nature of these events, the proposed method maintains bounded mechanical energy throughout the simulation, even in the absence of explicit viscous effects.
One counterintuitive observation is that the viscous simulations appear less dissipative than their inviscid counterparts.
The presence of viscosity continuously moderates the flow field, degrading the strength of the most violent events and inhibiting the transition toward turbulence.
As a result, the viscous flow does not experience dissipation with turbulence-like mechanisms typically associated with iLES, where finer-scale structures are permitted to develop and contribute to enhanced dissipation.
In contrast, the inviscid simulations allow richer small-scale dynamics through nonlinear energy transfer, which in turn lead to larger numerical dissipation.

Grid convergence of the mechanical energy evolution is shown in \cref{fig:sinewave_ECompare}.
Both the total energy and the dissipation rate exhibit consistent trends as the grid resolution is increased.
Notably, the dissipation rate does not vanish in the limit of infinite resolution.
This behavior can be attributed to the continuous decrease of the numerical Hinze and Kolmogorov scales, which enables increasingly fine, fractal-like fragmentation of the interface.
Combined with interface-localized upwinding, this mechanism yields a physically consistent, energy-dissipative limit.
In essence, the flow behaves as though it never reaches a scale below the Kolmogorov scale, instead sustaining a persistent cascading inertial range across all resolved scales even with increasing resolution.

\section{Conclusions}

This effort proposes modifications to existing advection and viscous schemes to enhance the fidelity and robustness of multiphase flow simulations. 
We provide both analytical and numerical demonstrations that most variants of Consistent Mass-Momentum (CMOM) methods violate the Total Variation Diminishing (TVD) condition near emptying interface cells, a deficiency that inherently threatens simulation credibility. 
Our analysis identifies momentum reconstruction and a mass-flux-based donor concept as necessary requirements to preserve exact TVD properties in multiphase flows. 
To address these needs, we developed a TVD \textit{Synchronized Donor Region of Momentum flux} (SynDRoM) for the CMOM framework, which significantly improves accuracy in predicting complex flow fields. 
The resulting algorithm, when combined with a directionally-split momentum update and a density-weighted RK2 scheme, robustly maintains TVD in multidimensional transport and handles extreme density ratios while retaining momentum conservation and empirical energy boundedness. 
These physical properties are validated through the Kelvin-Helmholtz instability.
We further introduce a bounded kinetic viscosity approach to handle the non-synchronicity of viscosity and density jump estimations, as demonstrated in falling film simulations. 
Both methods were further tested in energetic standing-wave breaker simulations and demonstrated the intended physical behaviors.

While this effort utilizes the simplest possible framework to demonstrate these core concepts, extensions to more advanced numerical setups are readily achievable. 
For the modified CMOM, the assumption of uniform density in the staggered grid can be relaxed, provided the total momentum in a cell is retained during reconstruction. 
Extension to unsplit schemes would require determining the momentum distribution where advection sweeps overlap and applying, for example, second-order Gaussian quadrature for integrating momentum fluxes. 
On the other hand, extension to collocated grids is relatively straightforward because the volume advection sweep is then identical to the mass donor region, eliminating the need for interpolation from collocated to staggered mass \citep{2013LeChenadec_Monotonicity}, but at the expense of more sophisticated velocity-pressure coupling, such as \citet{2025Santos_ProjectionCollocated}. 
The current methods also demonstrate controlled dissipation during high kinetic energy concentration events--such as the interaction of spilling breakers with flat interfaces and interface roll-ups during nonlinear instability phases. 
As noted in recent research, this proposed strategy can be readily applied to passive scalar transport \citep{2023Kuhn_MMC,2023Remmerswaal_Thesis} in scenarios such as chemical reactions, thixotropic fluids, and heat transport. 
Finally, the bounded kinetic viscosity approach is not restricted to CMOM and can be applied to standard velocity formulations. 
The base dynamic viscosity can also be further generalized using setups such as \citet{2002Prosperetti_Viscosity}, and the methodology remains compatible with both sharp-interface and diffused-interface methods.

% The proposed numerical schemes, together with conservative geometric Volume-of-Fluid method, are naturally compatible with topologically convoluted and changing interfaces.

The proposed numerical scheme is orthogonal to the interface capturing method, improving physical stability, and thus reliability, to the simulation. While topological changes are handled by the interface capturing scheme of choice (in our case, a conservative geometric Volume-of-Fluid), the enhanced physical reliability provided by SynDRoM and the viscosity limiter helps in removing numerical artifacts that could subsequently trigger false break-up events, always subject to an adequate mesh resolution. 
These characteristics make the approach particularly attractive for energetic free-surface and multiphase flows encountered in naval hydrodynamics.

% \clearpage

\section{Acknowledgments}
This work was fully supported by Delft University of Technology.

\bibliographystyle{ieeetrannAST}
\bibliography{references.bib}

\clearpage

\newcommand{\disspace}{\vspace{0.4cm}}
\newcommand{\discolor}{\color{teal}}

\section*{Discussions}
The authors want to thank the referees for dedicating time to review this work and providing a constructive feedback.

Comments (highlighted in teal) by External Discussers (ED), the Organizing and Papers Committee (OPC), and On-site Discussers (OD) are addressed below.

\subsection*{ED: Prof.~Pablo M.~Carrica}
General Dynamics Electric Boat, 50 Pequot Ave, New London, CT 06320, USA.

{\discolor
The authors present clever improvements to algorithms to compute the mass and momentum fluxes and to evaluate viscosity jumps to improve stability in large density and viscosity ratio flows, known to be very challenging.

1) The paper presents the problems of CMOM and how SynDRoM combined with density weighted RK2 and directionally-split momentum update can alleviate these problems. There is an interesting grid convergence study, which shows that finer grids resolve smaller bubbles and dissipation scales, but not reaching the Kolmogorov scale. Can the authors show that with some numbers? }

Inviscid formulations lack a physical lower bound on spatial length scales of flows and interfaces. 
As a result, the convergence behaves differently from the viscous simulation.
However, within the implicit Large-Eddy Simulation (iLES) framework \citepdiscussion{2008Aspden_iLES,2019Hendrickson_Stern}, numerical dissipation introduces a grid-restrained numerical Kolmogorov scale and interface scale. 

\begin{figure}[ht]
    \centering
    \includegraphics[width=0.95\linewidth]{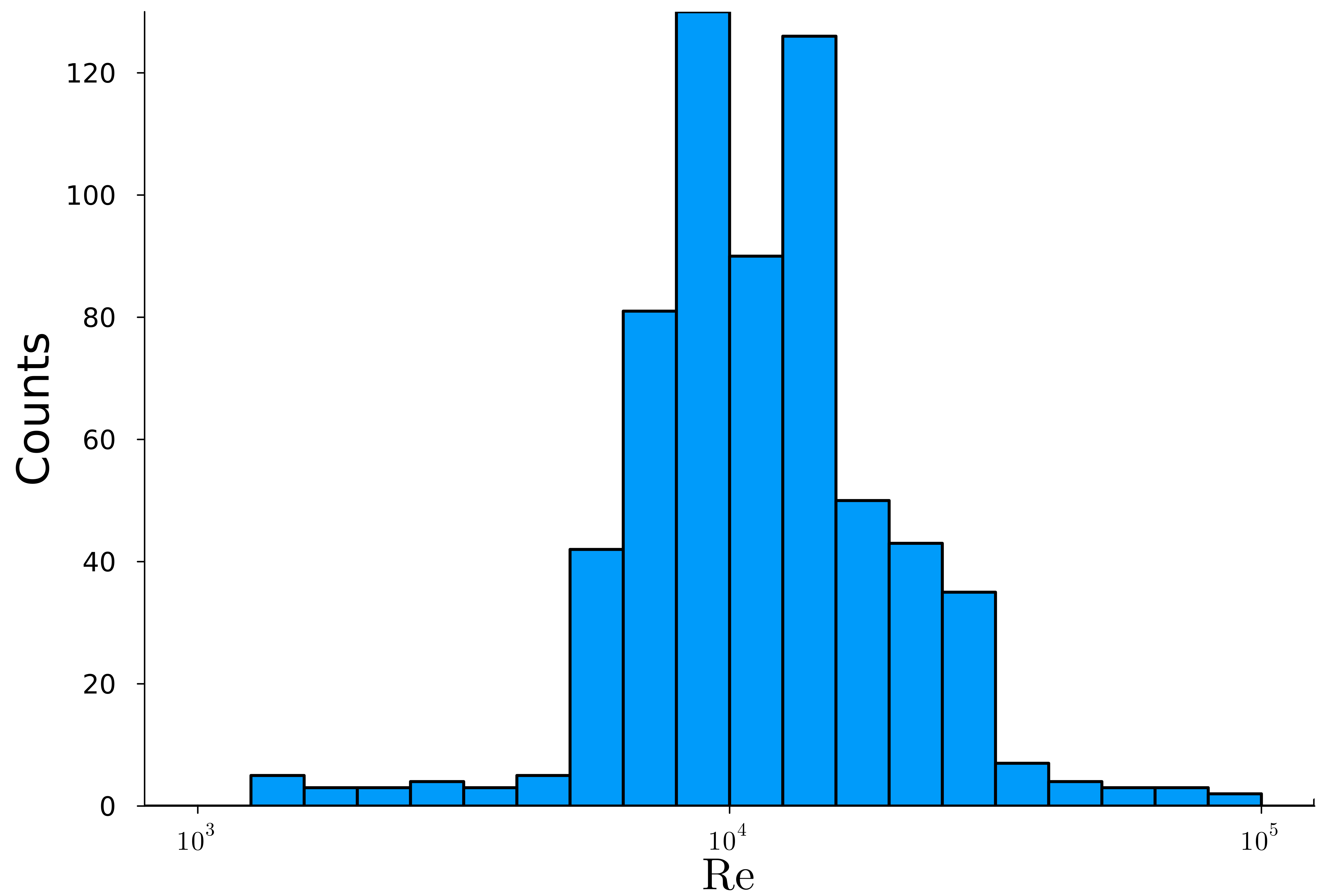}
    \caption{Histogram of effective Reynolds numbers $Re_\mathrm{eff}$ evaluated across time steps for the inviscid standing wave simulation ($N_x=N_y=N_z=256$).}
    \label{fig:ReHist}
\end{figure}

The numerical Kolmogorov scale can be quantified by defining an effective dynamic viscosity, $\mu_\mathrm{eff}$, derived from the global kinetic energy dissipation rate and the integrated strain-rate tensor in the water region $\Omega_\mathrm{w}$, assuming the energy exchange between air and water is negligible to the water dissipation:
\begin{align}
    \mu_\mathrm{eff} = \frac{-\partial_t \int_{\Omega_\mathrm{w}} \frac{1}{2}\rho u_i u_i \dd{V}}{2 \int_{\Omega_\mathrm{w}} S_{ij} S_{ij} \dd{V}}, \quad S_{ij} = \frac{1}{2}(\partial_i u_j + \partial_j u_i).
\end{align}
\Cref{fig:ReHist} shows the temporal distribution of the resulting effective Reynolds number, $Re_\mathrm{eff} = \rho_\mathrm{w} U L / \mu_\mathrm{eff}$, for the inviscid standing wave simulation on the finest grid ($N_x=N_y=N_z=256$). The bulk of the effective Reynolds numbers cluster around $10^4$, resulting in the effective Kolmogorov scale of $\mathcal{O}(10^{-3})$ \citepdiscussion{pope2000turbulent}, similar to the grid size $(1/256)$.

\disspace

{\discolor
2) While it is clear from the paper that the techniques presented work well for staggered grids, the authors claim that “extension to collocated grids is relatively straightforward.” However, pressure-velocity coupling and volume conservation problems that affect collocated methods would remain, preventing the desired improvements in stability. Please elaborate.}

The reviewer raises an important point regarding flow solver stability on collocated grids.
Our statement in the manuscript that the extension of SynDRoM to collocated grids is ``relatively straightforward'' referred specifically to the spatial synchronization of mass and momentum advection fluxes: on a collocated arrangement, the advection sweep volumes for mass and momentum naturally coincide, which eliminates the need for isometric averaging between staggered and collocated control volumes.

However, overall numerical robustness on collocated grids indeed requires dedicated treatment of pressure-velocity coupling, and is out of scope of this study. 
Standard collocated formulations are prone to high-frequency pressure-velocity decoupling and nonphysical checkerboard phenomenon. 
Symmetry-preserving collocated projection schemes should be applied -- such as that of \citetdiscussion{2025Santos_ProjectionCollocated} -- which explicitly preserve discrete differential identities (\textit{e.g.}, ensuring zero divergence of the discrete pressure gradient).
The modification is still straightforward as it mainly modified the interpolation operators.

We have revised the Concluding Remarks to clarify that synchronization between mass and momentum in collocated meshes is straightforward, albeit the pressure-velocity coupling needs cares.

\disspace

{\discolor
3) And how about independence to time step? Is SynDRoM and associated algorithms well behaved for time steps from required by CFL all the way down to very small time steps?
}

% SynDRoM is stable even if the time step size is small. 
% However, SynDRoM might have some issues when the time step size is comparable to the divergence tolerance (pressure solver tolerance) as the SynDRoM or CMOM-based method requires discrete mass conservation.
% Currently the CFL in 3D cVOF algorithm is around $0.1$ \citepdiscussion{2010Weymouth_cVOF}, and the pressure solver tolerance is $10^{-4}$ for single precision. 

% However, in our previous research the spatial error such as reconstruction and directional-split generally overwhelm the temporal error and introduce the so-called numerical shape diffusion effect \citepdiscussion{2010Weymouth_cVOF}.
% As a result, we generally avoid small time step size since it brings more steps of reconstruction and amplify the numerical shape diffusion effect. 

SynDRoM and its associated integration algorithms remain robust and bounded across the full spectrum of time steps, from the maximum allowable CFL condition down to the small time-step limit ($\mathrm{CFL} \to 0$). 
The underlying TVD flux-limiting framework satisfies velocity monotonicity independently of CFL.

However, reducing CFL far below the limit brings spatial error accumulation, a phenomenon we coined as numerical shape diffusion \citepdiscussion{2024Huang_NSH}. 
Taking excessively small time steps increases the number of geometric reconstruction and splitting operations per physical time unit, thereby accelerating numerical shape diffusion of the interface.
Within directionally split cVOF frameworks, spatial discretization errors, specifically interface reconstruction and operator splitting, dominate temporal integration errors. 

% However, two numerical considerations govern behavior at extremely small CFL:

% \begin{itemize}
%     \item \textbf{Pressure Solver Residual Coupling:} 
%     Exact mass-momentum synchronization in SynDRoM relies on discrete mass conservation to certain tolerance. 
%     The operational time step must therefore remain larger than the residual tolerance of the pressure Poisson solver ($\epsilon_\mathrm{tol}$). 
%     If CFL approaches $O(\epsilon_\mathrm{tol})$, where typical single-precision solver tolerances are $\epsilon_\mathrm{tol} \approx 10^{-4}$, unreduced divergence residuals can impair the exact alignment of mass and momentum fluxes. 
%     Standard operational time steps set by the 3D cVOF limit ($\mathrm{CFL} \approx 0.2$) \citepdiscussion{2010Weymouth_cVOF} maintain a sufficient margin.
%     \item \textbf{Accumulation of Numerical Shape Diffusion:} Reducing CFL far below the limit is computationally counterproductive due to spatial error accumulation. 
%     Taking excessively small time steps increases the number of geometric reconstruction and splitting operations per physical time unit, thereby accelerating numerical shape diffusion of the interface.
%     Within directionally split cVOF frameworks, spatial discretization errors, specifically interface reconstruction and operator splitting, dominate temporal integration errors \citepdiscussion{2024Huang_NSH}. 
% \end{itemize}

\subsection*{OPC}
{\discolor
The Synchronized Donor-Region of Momentum fluxes (Syn-DRoM) method appears to offer a robust solution to the TVD (total variation diminishing) violation in CMOM (Consistent Mass-Momentum) schemes. Could the authors comment on the computational overhead introduced by the synchronized donor region approach, especially in large-scale 3D simulations?
}

Evaluating momentum fluxes with SynDRoM increases the floating-point operation (FLOP) count of the advection step by roughly a factor of one compared to that of the standard slope limiters (\textit{e.g.}, the Koren limiter). 
However, since momentum advection accounts for roughly 5\% of the total runtime per time step which is mostly dominated by the pressure solver, this localized FLOP increase has minor impact on overall execution time.

Crucially, this minor local overhead is offset by computational savings in the pressure projection step, which is the primary computational bottleneck in 3D incompressible multiphase solvers. 
By suppressing nonphysical high-frequency velocity field near high-density-ratio interfaces, SynDRoM  improves the solvability (effective condition number) of the discrete system, thereby reducing the iteration count required by the pressure Poisson solver to reach convergence. 
Consequently, the net wall-clock time per time step for SynDRoM in large-scale 3D simulations is comparable to, or in practice lower than, standard CMOM or traditional velocity schemes.

We have revised the Introduction section of the manuscript to explicitly address this computational trade-off.

\disspace

{\discolor
In the context of interface-capturing vs. interface-reconstructing methods, do the authors foresee any limitations of Syn-DRoM when applied to flows with highly convoluted or topologically changing interfaces?
}

SynDRoM method is independent of any interface advection method.
However, by preventing air-phase velocity from penetrating into the dense phase, SynDRoM, or CMOM generally, eliminates nonphysical interface distortion (\cref{fig:dambreak_shape}a), yielding sharper interface topology than traditional velocity formulations \citepdiscussion{2022Liu_CMOM}.
The method is broadly compatible with both diffuse interface-capturing schemes using indicator functions (\textit{e.g.}, phase-field, level-set) and sharp interface-reconstructing methods such as geometric Volume-of-Fluid (gVOF) \citepdiscussion{2017Mirjalili_IC}. 
Sharp interface-reconstructing approaches achieve accuracy comparable to diffuse models at coarser grid resolutions, and meanwhile SynDRoM ensures these sharp interfaces remain numerically robust during complex topological changes.

\disspace

{\discolor
The results for the 3D standing wave breaker are impressive. Could the authors elaborate on the scalability of their implementation and its applicability to real-world naval hydrodynamics problems, such as ship-wave interactions or propeller cavitation?
}

Our multiphase solver is under development to support multi-GPU execution via the Message Passing Interface (MPI), enabling scaling to large 3D naval hydrodynamics applications given adequate GPU capacity \citepdiscussion{2026Font_MPI}. 

Regarding real-world applicability, the framework is directly suited for ship-wave interactions. 
Ongoing developments focus on integrating body-immersion methods for fluid-structure interaction and non-reflecting wave radiation boundary conditions to minimize computational domain size. 
Conversely, phase-change phenomena such as propeller cavitation are outside the immediate scope of current work, though the modular Julia ecosystem facilitates future incorporation of cavitation physics models.

\disspace

{\discolor
Finally, the open-source release of InterfaceAdvection.jl is commendable. Are there plans to integrate these methods into widely used CFD platforms, or to provide benchmarks for community validation?
}

To facilitate community validation, we are developing dedicated benchmark repositories for canonical problems in naval hydrodynamics. 
For example, \texttt{UnderwaterTurbulence.jl}\footnotemark{} provides reproducible setups for free-surface turbulent flow benchmarks \citepdiscussion{2010Guo_HITFreeSurface,2025Calado_HIT_FreeSurface,2025Gaylo_FST_HIT}.

We do not plan to integrate these schemes into external CFD suites however. 
\texttt{InterfaceAdvection.jl} is intended as a standalone framework, and maintaining a self-contained Julia codebase allows us to preserve optimal computational performance and flexibility.

\footnotetext{\url{https://github.com/TzuYaoHuang/UnderwaterTurbulence.jl}}

\disspace

{\discolor
A minor editorial suggestion: In the title page, close the affiliation of the authors by adding a closing parenthesis “)”.
}
Fixed.

\subsection*{OD: Prof.~Xiaofeng Liu}
Department of Aerospace Engineering, San Diego State University, SD, USA.

{\discolor
Hi Tzu-Yao, Thanks for the presentation. But talking about momentum balance by completely neglecting the viscous term as shown in your momentum equation on your Side 7 does not make sense because you neglected the very fundamental physics that viscosity plays during the sharp-interface evolving process. Neglecting the viscous effect is completely wrong if you want to get the modeling right.
}

We fully agree that viscous stress plays an important physical role in real-world sharp-interface dynamics. 
However, omitting the viscous term in the initial advection formulation is an intentional methodological choice to isolate the behavior of the advection term.

Physical viscosity introduces real dissipation that can artificially mask purely numerical instabilities, such as spurious momentum leakage, non-monotonic velocity overshoots, or kinetic energy divergence. 
Evaluating the advection operator in the inviscid limit forces the scheme to prove its intrinsic TVD properties and energy-boundedness solely through mass-momentum synchronization, ensuring that numerical stability is not falsely attributed to physical damping.

Having established the intrinsic robustness of the advection framework, the second part of our work explicitly addresses interfacial viscous mechanics. 
We introduce a bounded kinetic viscosity limiter to handle severe viscosity jumps while maintaining viscous CFL stability, and demonstrate the complete formulation on viscous benchmarks (e.g., standing wave breakers).

\bibliographystylediscussion{ieeetrannAST}
\bibliographydiscussion{references-discussion.bib}

% \end{multicols*}
\end{document}